\documentclass[journal]{IEEEtran}

\usepackage{cite}
\ifCLASSINFOpdf
\else
\fi
\usepackage{amsmath}

\DeclareUnicodeCharacter{2061}{}

\usepackage{graphicx}
\usepackage{balance}
\usepackage{hyperref}
\hypersetup{
    colorlinks=true,
    linkcolor=black,
    filecolor=black,      
    urlcolor=blue,
    citecolor=black,
    pdfpagemode=FullScreen,
}

\usepackage[dvipsnames]{xcolor}
\begin{document}
\title{Deep Reinforcement Learning for Cyber Security}

\author{Thanh~Thi~Nguyen~%,~\IEEEmembership{Member,~IEEE,}
        %John~Doe,~\IEEEmembership{Fellow,~OSA,}
        and~Vijay~Janapa~Reddi%,~\IEEEmembership{Life~Fellow,~IEEE}% <-this % stops a space
\thanks{T. T. Nguyen is with the School of Information Technology, Deakin University, Melbourne Burwood Campus, Burwood, VIC 3125, Australia, e-mail: thanh.nguyen@deakin.edu.au.}% <-this % stops a space
\thanks{V. J. Reddi is with the John A. Paulson School of Engineering and Applied Sciences, Harvard University, Cambridge, MA 02138, USA, e-mail: vj@eecs.harvard.edu.}% <-this % stops a space
%\thanks{Manuscript received June 19, 2019; revised June 26, 2019.}
}


\markboth{IEEE Transactions on Neural Networks and Learning Systems, \href{https://doi.org/10.1109/TNNLS.2021.3121870}{DOI: 10.1109/TNNLS.2021.3121870}, Early Access}% https://doi.org/10.1109/TNNLS.2021.3121870
{Nguyen and Reddi: Deep Reinforcement Learning for Cyber Security}




\maketitle

\begin{abstract}
The scale of Internet-connected systems has increased considerably, and these systems are being exposed to cyber attacks more than ever. The complexity and dynamics of cyber attacks require protecting mechanisms to be responsive, adaptive, and scalable. Machine learning, or more specifically deep reinforcement learning (DRL), methods have been proposed widely to address these issues. By incorporating deep learning into traditional RL, DRL is highly capable of solving complex, dynamic, and especially high-dimensional cyber defense problems. This paper presents a survey of DRL approaches developed for cyber security. We touch on different vital aspects, including DRL-based security methods for cyber-physical systems, autonomous intrusion detection techniques, and multiagent DRL-based game theory simulations for defense strategies against cyber attacks. Extensive discussions and future research directions on DRL-based cyber security are also given. We expect that this comprehensive review provides the foundations for and facilitates future studies on exploring the potential of emerging DRL to cope with increasingly complex cyber security problems.
\end{abstract}


% \fi
\IEEEpeerreviewmaketitle

\section{Introduction}
\IEEEPARstart{I}{nternet} of Things (IoT) technologies have been employed broadly in many sectors such as telecommunications, transportation, manufacturing, water and power management, healthcare, education, finance, government, and even entertainment. The convergence of various information and communication technology (ICT) tools in the IoT has boosted its functionalities and services to users to new levels. ICT has witnessed a remarkable development in terms of system design, network architecture, and intelligent devices in the last decade. For example, ICT has been advanced with the innovations of cognitive radio network and 5G cellular network \cite{Kakalou2017, Huang2020}, software-defined network (SDN) \cite{Wang2020}, cloud computing \cite{Botta2016}, (mobile) edge caching \cite{Krestinskaya2020, Abbas2018}, and fog computing \cite{Dastjerdi2016}. Accompanying these developments is the increasing vulnerability to cyber attacks, which are defined as any type of offensive maneuver exercised by one or multiple computers to target computer information systems, network infrastructures, or personal computer devices. Cyber attacks may be instigated by economic competitors or state-sponsored attackers. There has been thus a critical need of the development of cyber security technologies to mitigate and eliminate impacts of these attacks \cite{Geluvaraj2019}.\par

Artificial intelligence (AI), especially machine learning (ML), has been applied to both attacking and defending in the cyberspace. On the attacker side, ML is utilized to compromise defense strategies. On the cyber security side, ML is employed to put up robust resistance against security threats in order to adaptively prevent and minimise the impacts or damages occurred. Among these ML applications, unsupervised and supervised learning methods have been used widely for intrusion detection \cite{Buczak2016, Apruzzese2018, Xin2018}, malware detection \cite{Milosevic2017, Mohammed2018, Berman2019}, cyber-physical attacks \cite{Paul2020, Ding2018, Wu2019}, and data privacy protection \cite{Xiao2018a}. In principle, unsupervised methods explore the structure and patterns of data without using their labels while supervised methods learn by examples based on data's labels. These methods, however, cannot provide \textit{dynamic and sequential responses} against cyber attacks, especially new or constantly evolving threats. Also, the detection and defending responses often take place after the attacks when traces of attacks become available for collecting and analyzing, and thus proactive defense solutions are hindered. A statistical study shows that 62\% of the attacks were recognized after they have caused significant damages to the cyber systems \cite{Sharma2011}.\par

Reinforcement learning (RL), a branch of ML, is the closest form of human learning because it can learn by its own experience through exploring and exploiting the unknown environment. RL can model an autonomous agent to take sequential actions optimally without or with limited prior knowledge of the environment, and thus, it is particularly adaptable and useful in real time and adversarial environments. With the power of \textit{function approximation} and \textit{representation learning}, deep learning has been incorporated into RL methods and enabled them to solve many complex problems \cite{Nguyen2017, Sui2020, Tsantekidis2020, Nguyen2018b, Wang2020b}. The combination of deep learning and RL therefore indicates excellent suitability for cyber security applications where cyber attacks are increasingly sophisticated, rapid, and ubiquitous \cite{Ling2015, Wang2019a, Lu2020, Alauthman2020}.\par

The emergence of DRL has actually witnessed great success in different fields, from video game domain, e.g. Atari \cite{Mnih2015, Nguyen2018a}, game of Go \cite{Silver2016, Silver2017}, real-time strategy game StarCraft II \cite{Vinyals2017, Sun2018a, Pang2018, Zambaldi2018}, 3D multi-player game Quake III Arena Capture the Flag \cite{Jaderberg2018}, and teamwork game Dota 2 \cite{OpenAI2019} to real-world applications such as robotics \cite{Gu2017}, autonomous vehicles \cite{Isele2018}, autonomous surgery \cite{Nguyen2019a, Nguyen2019b}, natural language processing \cite{Keneshloo2020}, biological data mining \cite{Mahmud2018}, and drug design \cite{Popova2018}. DRL methods have also recently been applied to solve various problems in the IoT area. For example, a DRL-based resource allocation framework that integrates networking, caching, and computing capabilities for smart city applications is proposed in \cite{He2017}. DRL algorithm, i.e., double dueling deep Q-network \cite{Hasselt2016, Wang2016a}, is used to solve this problem because it involves a large state space, which consists of the dynamic changing status of base stations, mobile edge caching (MEC) servers and caches. The framework is developed based on the programmable control principle of SDN and the caching capability of \textit{information-centric networking}. Alternatively, Zhu et al. \cite{Zhu2018a} explored MEC policies by using the \textit{context awareness} concept that represents the user's context information and traffic pattern statistics. The use of AI technologies at the mobile network edges is advocated to intelligently exploit operating environment and make the right decisions regarding what, where, and how to cache appropriate contents. To increase the caching performance, a DRL approach, i.e., the asynchronous advantage actor-critic algorithm \cite{Mnih2016}, is used to find an optimal policy aiming to maximize the offloading traffic. \par

Findings from our current survey show that applications of DRL in cyber environments are generally categorized under two perspectives: optimizing and enhancing the communications and networking capabilities of the IoT applications, e.g. \cite{Zhang20171, Zhu20181, Shafin2020, Zhang2018, He20181, He20182, Luong2019, Dai2019, Leong2020}, and defending against cyber attacks. This paper focuses on the later where DRL methods are used to solve cyber security problems with the presence of cyber attacks or threats. Next section provides a background of DRL methods, followed by a detailed survey of DRL applications in cyber security in Section III. We group these applications into three major categories, including DRL-based security solutions for cyber-physical systems, autonomous intrusion detection techniques, and DRL-based game theory for cyber security. Section IV concludes the paper with extensive discussions and future research directions on DRL for cyber security.

\section{Deep Reinforcement Learning Preliminary}
\begin{figure}[htp]
\centering
\includegraphics[width=0.29\textwidth]{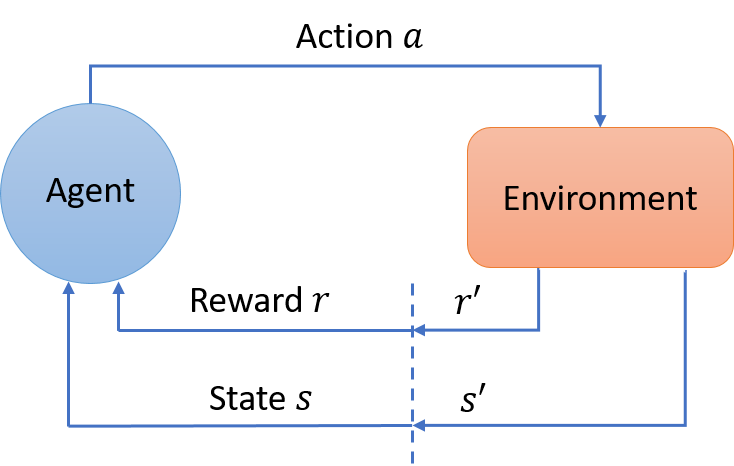}
\caption{Interactions between the agent and its environment in RL, characterized by state, action and reward. Based on the current state $s$ and reward $r$, the agent will take an optimal action, leading to changes of state and reward. The agent then receives the next state $s'$ and reward $r'$ from the environment to determine the next action, making an iterative process of agent-environment interactions.}
\label{fig:1}
\end{figure}

Different from the other popular branch of ML, i.e., supervised methods learning by examples, RL characterizes an \textit{agent} by creating its own learning experiences through interacting directly with the \textit{environment}. RL is described by concepts of \textit{state}, \textit{action}, and \textit{reward} (Fig. \ref{fig:1}). It is a trial and error approach in which the agent takes action at each time step that causes two changes: current state of the environment is changed to a new state, and the agent receives a reward or penalty from the environment. Given a state, the reward is a function that can tell the agent how good or bad an action is. Based on received rewards, the agent learns to take more good actions and gradually filter out bad actions.

A popular RL method is Q-learning whose goal is to maximize the discounted cumulative reward based on the Bellman equation \cite{Watkins1992}:
\begin{equation} \label{eq1}
Q(s_t,a_t)=E[r_{t+1}+\gamma r_{t+2}+\gamma^2 r_{t+3}+...|s_t,a_t]
\end{equation}
The discount factor \(\gamma\in[0,1]\) manages the importance levels of future rewards. It is applied as a mathematical trick to analyze the learning convergence. In practice, discount is necessary because of partial observability or uncertainty of the stochastic environment.\par

Q-learning needs to use a lookup table or Q-table to store expected rewards (Q-values) of actions given a set of states. This requires a large memory when the state and action spaces increase. Real-world problems often involve continuous state or action space, and therefore, Q-learning is inefficient to solve these problems. Fortunately, deep learning has emerged as a powerful tool that is a great complement to traditional RL techniques. Deep learning methods have two typical capabilities, i.e. function approximation and representation learning, which help them to learn a compact low-dimensional representation of raw high-dimensional data effectively \cite{Arulkumaran2017}. The combination of deep learning and RL was the research direction that Google DeepMind has initiated and pioneered. They proposed deep Q-network (DQN) with the use of a deep neural network (DNN) to enable Q-learning to deal with high-dimensional sensory inputs \cite{Mnih2013, Mnih2015}.\par

\begin{figure}[bp]
\centering
\includegraphics[width=0.49\textwidth]{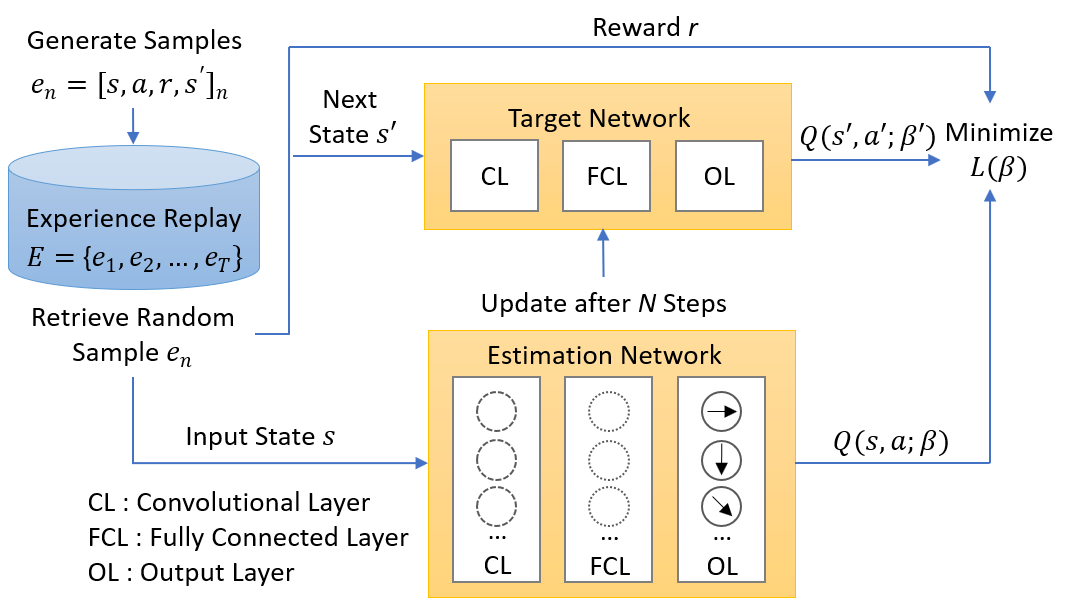}
\caption{DQN architecture with the loss function described by \(L(\beta)=E[(r+\gamma \max_{a'}⁡Q(s',a'|\beta')-Q(s,a|\beta))^2]\) where \(\beta\) and \(\beta'\) are parameters of the estimation and target deep neural networks respectively. Each action taken by the agent will generate an experience, which consists of the current state \(s\), action \(a\), reward \(r\) and next state \(s'\). These learning experiences (samples) are stored in the experience replay memory, which are then retrieved randomly for a stable learning process.}
\label{fig:2}
\end{figure}

Using DNNs to approximate the Q-function however is unstable due to the correlations among the sequences of observations and the correlations between the Q-values \(Q(s,a)\) and the target values \(Q(s',a')\). Mnih et al. \cite{Mnih2015} proposed the use of two novel techniques, i.e., experience replay memory and target network, to address this problem (Fig. \ref{fig:2}). On the one hand, experience memory stores an extensive list of learning experience tuples \((s,a,r,s')\), which are obtained from the agent's interactions with the environment. The agent's learning process retrieves these experiences randomly to avoid the correlations of consecutive experiences. On the other hand, the target network is technically a copy of the estimation network, but its parameters are frozen and only updated after a period. For instance, the target network is updated after 10,000 updates of the estimation network, as demonstrated in \cite{Mnih2015}. DQN has made a breakthrough as it is the first time an RL agent can provide a human-level performance in playing 49 Atari games by using just raw image pixels of the game board.\par

\begin{figure}[htp]
\centering
\includegraphics[width=0.38\textwidth]{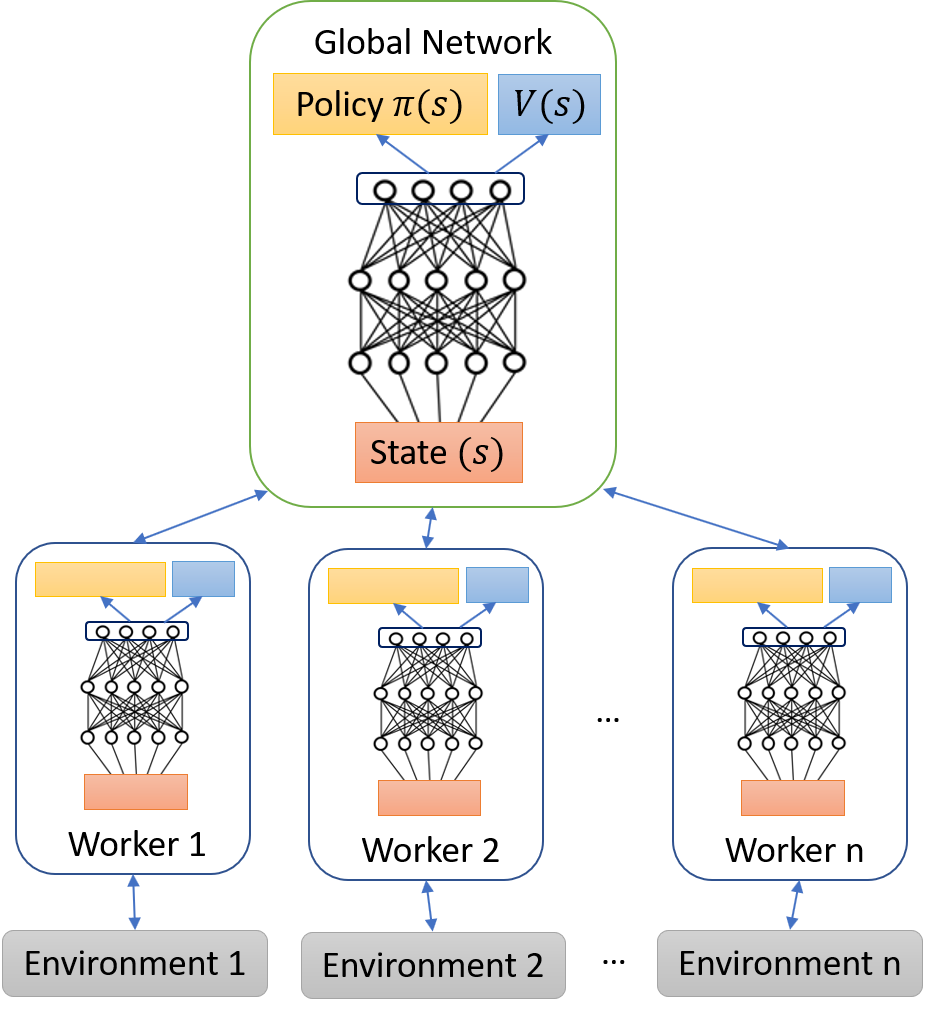}
\caption{The learning architecture of A3C, consisting of a global network and a number of worker agents. Each worker initially resets its parameters to those of the global network and interacts with its copy of the environment for learning. Gradients obtained from these individual learning processes will be used to update the global network asynchronously. This increases the learning speed and diversifies the experience learned by the global network as the experiences obtained by individual worker agents are independent.}
\label{fig:3}
\end{figure}

As a \textit{value-based method}, DQN takes long training time and has limitations in solving problems with continuous action spaces. Value-based methods, in general, evaluate the goodness of an action given a state using the Q-value function. When the number of states or actions is large or infinite, they show inefficiency or even impracticality. Another type of RL, i.e., \textit{policy gradient methods}, has solved this problem effectively. These methods aim to derive actions directly by learning a policy \(\pi(s,a)\) that is a probability distribution over all possible actions. REINFORCE \cite{Williams1992}, vanilla policy gradient \cite{Sutton2000}, trust region policy optimization (TRPO) \cite{Schulman2015} and proximal policy optimization (PPO) \cite{Schulman2017} are notable policy gradient methods. The gradient estimation, however, often suffers from a large fluctuation \cite{Wu2018}. The combination of value-based and policy-gradient methods has been developed to aggregate the advantages and eradicate the disadvantages of these two methods. This kind of combination has constituted another type of RL, i.e., \textit{actor-critic methods}. This structure comprises two components: an actor and a critic that can be both characterized by DNNs. The actor attempts to learn a policy by receiving feedback from the critic. This iterative process helps the actor improve its strategy and converge to an optimal policy. Deep deterministic policy gradient (DDPG) \cite{Lillicrap2015}, distributed distributional DDPG (D4PG) \cite{Barth-Maron2018}, asynchronous advantage actor-critic (A3C) \cite{Mnih2016} and unsupervised reinforcement and auxiliary learning (UNREAL) \cite{Jaderberg2016} are methods that utilize the actor-critic framework. An illustrative architecture of the popular algorithm A3C is presented in Fig. \ref{fig:3}. A3C's structure consists of a hierarchy of a master learning agent (global) and individual learners (workers). Both master agent and individual learners are modeled by DNNs with each having two outputs: one for the critic and another for the actor. The first output is a scalar value representing the expected reward of a given state $V(s)$ while the second output is a vector of values representing a probability distribution over all possible actions $\pi(s,a)$.

The value loss function of the critic is specified by:
\begin{equation} \label{eq2}
L_1=\sum(R-V(s))^2  
\end{equation}
where \(R=r+\gamma V(s')\) is the discounted future reward. Also, the actor is pursuing minimization of the following policy loss function:
\begin{equation} \label{eq3}
L_2=-\log(\pi(a|s))*A(s)-\vartheta H(\pi)
\end{equation}
where \(A(s)=R-V(s)\) is the estimated \textit{advantage} function, and \(H(\pi)\) is the entropy term, which handles the exploration capability of the agent with the hyperparameter \(\vartheta\) controlling the strength of the entropy regularization. The advantage function \(A(s)\) shows how advantageous the agent is when it is in a particular state. The learning process of A3C is \textit{asynchronous} because each learner interacts with its separate environment and updates the master network independently. This process is iterated, and the master network is the one to use when the learning is finished.\par

\begin{table*}[ht]
\centering
\caption{Summary of features of DRL types and their notable methods}
\label{tb:1}
\begin{tabular}{|p{1.4cm}|p{5.7cm}|p{3.9cm}|p{4cm}|}
 \hline
DRL types & Value-based & Policy-gradient & Actor-critic\\ 
\hline
Features
& 
- Compute value of action given a state \(Q(s,a)\).\newline
- No learned explicit policy.\newline
- Sample efficient \cite{Nachum2017}.
&
- No value function is needed.\newline
- Explicit policy is constructed.\newline
- Sample inefficient \cite{Nachum2017}.
&
- Actor produces policy \(\pi(s,a)\).\newline
- Critic evaluates action by \(V(s)\).\newline
- Often perform better than value-based or policy-gradient methods.
\\
\hline
Typical methods
&
- DQN \cite{Mnih2015}\newline
- Double DQN \cite{Hasselt2016}\newline
- Dueling Q-network \cite{Wang2016a}\newline
- Prioritized Experience Replay DQN \cite{Schaul2015}
&
- REINFORCE \cite{Williams1992}\newline
- Vanilla Policy Gradient \cite{Sutton2000}\newline
- TRPO \cite{Schulman2015}\newline
- PPO \cite{Schulman2017}
&
- DDPG \cite{Lillicrap2015}\newline
- D4PG \cite{Barth-Maron2018}\newline
- A3C \cite{Mnih2016}\newline
- UNREAL \cite{Jaderberg2016}
\\
\hline
Applications
& 
Suitable for problems with discrete action spaces, e.g., classic control tasks: Acrobot, CartPole, and MountainCar as described and implemented in the popular OpenAI Gym toolkit \cite{OpenAI1}.
&
\multicolumn{2}{|p{7.9cm}|}{More suitable for problems with continuous action spaces, e.g., classic control tasks described and implemented in the OpenAI Gym toolkit: MountainCarContinuous and Pendulum \cite{OpenAI1} or BipedalWalker and CarRacing problems \cite{OpenAI2}.}
\\
\hline
\end{tabular}
\end{table*}

Table \ref{tb:1} summarizes comparable features of value-based, policy-gradient, and actor-critic methods, and their typical example algorithms. Valued-based methods are more sample efficient than policy-gradient methods because they are able to exploit data from other sources such as experts \cite{Nachum2017}. In DRL, a value function or a policy function is normally approximated by a universal function approximator such as a (deep) neural network, which can take either discrete or continuous states as inputs. Therefore, modelling state spaces is more straightforward than dealing with action spaces in DRL. Value-based methods are suitable for problems with discrete action spaces as they evaluate every action explicitly and choose an action at each time step based on these evaluations. On the other hand, policy-gradient and actor-critic methods are more suitable for continuous action spaces because they describe the policy (a mapping between states and actions) as a probability distribution over actions. The continuity characteristic is the main difference between discrete and continuous action spaces. In a discrete action space, actions are characterized as a mutually exclusive set of options while in a continuous action space, an action has a value from a certain range or boundary \cite{Lapan2018}.

\section{DRL in Cyber Security: A Survey}
A large number of applications of RL to various aspects of cyber security have been proposed in the literature, ranging from data privacy to critical infrastructure protection. However, drawbacks of traditional RL have restricted its capability in solving complex and large-scale cyber security problems. The increasing number of connected IoT devices in recent years have led to a significant increase in the number of cyber attack instances as well as their complexity. The emergence of deep learning and its integration with RL have created a class of DRL methods that are able to detect and fight against sophisticated types of cyber attacks, such as falsified data injection to cyber-physical systems \cite{Akazaki2018}, deception attack to autonomous systems \cite{Gupta2018}, distributed denial-of-service attacks \cite{Malialis2015}, intrusions to host computers or networks \cite{Lopez2020}, jamming \cite{Xiao2018b}, spoofing \cite{Xiao2016}, malware \cite{Wan2017}, attacks in adversarial networking environments \cite{Han2018}, and so on. This section provides a comprehensive survey of state-of-the-art DRL-powered solutions for cyber security, ranging from defense methods for cyber-physical systems to autonomous intrusion detection approaches, and game theory-based solutions.

\begin{figure}[htp]
\centering
\includegraphics[width=0.49\textwidth]{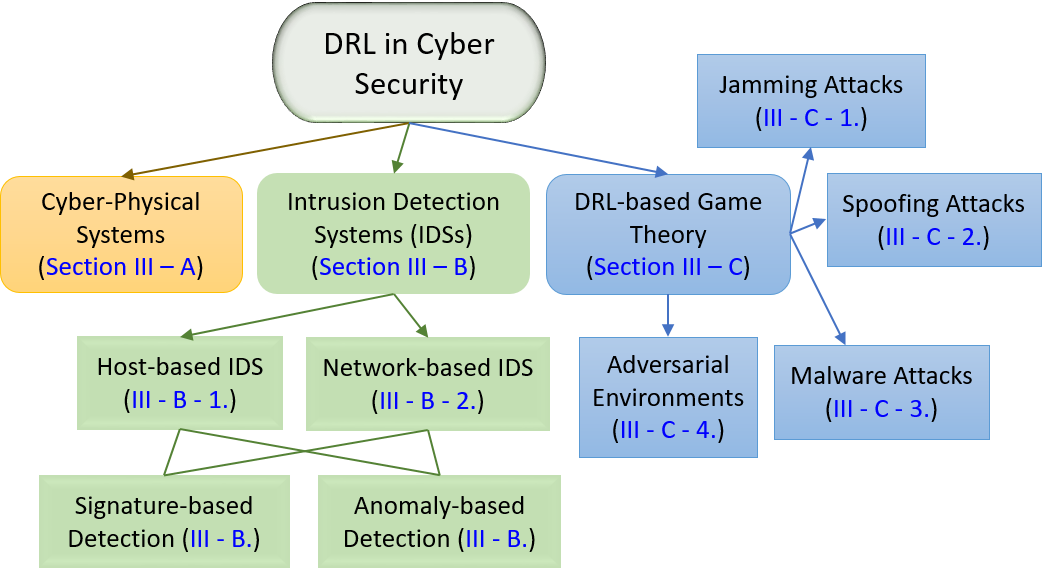}
\caption{Different (sub)sections of the survey on DRL in cyber security.}
\label{fig_structure}
\end{figure}

The structure of the survey is presented in Fig. \ref{fig_structure}. We limit the survey to existing applications of DRL to cyber security. There are other topics in cyber security where DRL has not been applied to and they are therefore discussed in Section IV (Discussions and Future Research Directions). Those potential topics include a multi-agent DRL approach to cyber security, combining host-based and network-based intrusion detection systems, model-based DRL and combining model-free and model-based DRL methods for cyber security applications, investigating methods that can deal with continuous action spaces in cyber environments, offensive AI, deepfakes, machine learning poisoning, adversarial machine learning, human-machine teaming within a human-on-the-loop architecture, bit-and-piece distributed denial-of-service attacks as well as potential attacks by quantum physics-based powerful computers to crack encryption algorithms.

\subsection{DRL-based Security Methods for Cyber-Physical Systems}
Investigations of defense methods for cyber-physical systems (CPS) against cyber attacks have received considerable attention and interests from the cyber security research community. CPS is a mechanism controlled by computer-based algorithms facilitated by internet integration. This mechanism provides efficient management of distributed physical systems via the shared network. With the rapid development of the Internet and control technologies, CPSs have been used extensively in many areas including manufacturing \cite{Wang2015a}, health monitoring \cite{Zhang2017, Shakeel2018}, smart grid \cite{Cintuglu2017, Chen2018, Ni2019}, and transportation \cite{Li2018, Ferdowsi2020}. Being exposed widely to the Internet, these systems are increasingly vulnerable to cyber attacks \cite{Li2019}. In 2015, hackers attacked the control system of a steel mill in Germany by obtaining login credentials via the use of phishing emails. This attack caused a partial plant shutdown and resulted in damage of millions of dollars. Likewise, there was a costly cyber attack to a power grid in Ukraine in late December 2015 that disrupted electricity supply to a few hundred thousand end consumers \cite{Feng2017}.\par

\begin{figure}[htp]
\centering
\includegraphics[width=0.35\textwidth]{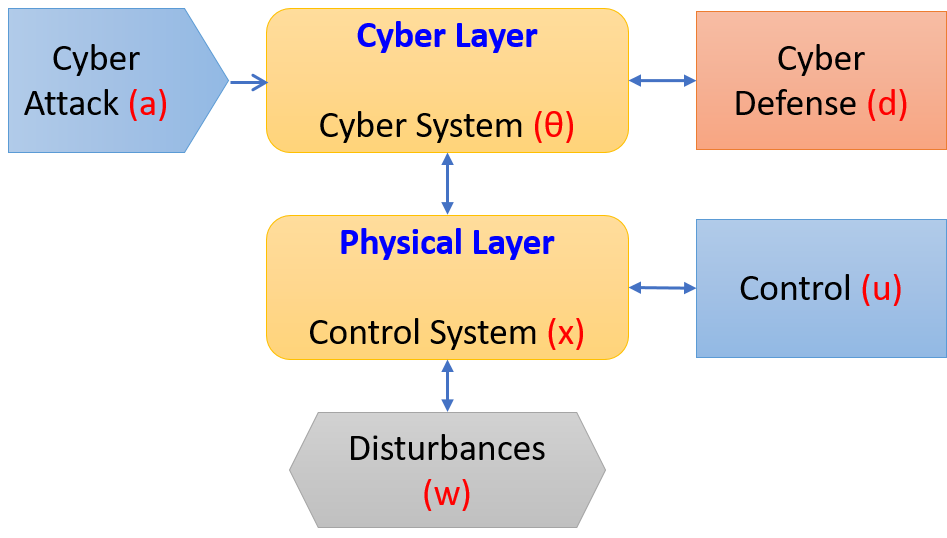}
\caption{The dynamics of attack and defense in a cyber-physical system. The physical layer is often uncertain with disturbances \(w\) while cyber attack \(a\) directly affects the cyber layer where a defense strategy \(d\) needs to be implemented. The dynamics of attack-defense characterized by \(\theta(t,a,d)\) is injected into the conventional physical system to develop a cyber-physical co-modelling as presented in Eq. (\ref{eq4}).}
\label{fig:5}
\end{figure}

In an effort to study cyber attacks on CPS, Feng et al. \cite{Feng2017} characterized the cyber state dynamics by a mathematical model:
\begin{equation} \label{eq4}
\dot{x}(t)=f(t,x,u,w;\theta(t,a,d)); \; \; \;  x(t_0 )=x_0
\end{equation}
where \(x\), \(u\) and \(w\) represent the physical state, control inputs and disturbances correspondingly (see Fig. \ref{fig:5}). In addition, \(\theta(t,a,d)\) describes cyber state at time \(t\) with \(a\) and \(d\) referring to cyber attack and defense respectively. \par

The CPS defense problem is then modeled as a two-player zero-sum game by which utilities of players are summed up to zero at each time step. The defender is represented by an actor-critic DRL algorithm. Simulation results demonstrate that the proposed method in \cite{Feng2017} can learn an optimal strategy to timely and accurately defend the CPS from unknown cyber attacks.\par

Applications of CPS in critical safety domains such as autonomous automotive, chemical process, automatic pilot avionics, and smart grid require a certain correctness level. Akazaki et al. \cite{Akazaki2018} proposed the use of DRL, i.e., double DQN and A3C algorithms, to find falsified inputs (counterexamples) for CPS models. This allows for effective yet automatic detection of CPS defects. Due to the infinite state space of CPS models, conventional methods such as simulated annealing \cite{Abbas2012} and cross entropy \cite{Sankaranarayanan2012} were found inefficient. Experimental results show the superiority of the use of DRL algorithms against those methods in terms of the smaller number of simulation runs. This leads to a more practical detection process for CPS models' defects despite the great complexity of CPS's software and physical systems.\par

\begin{figure}[bp]
\centering
\includegraphics[width=0.3\textwidth]{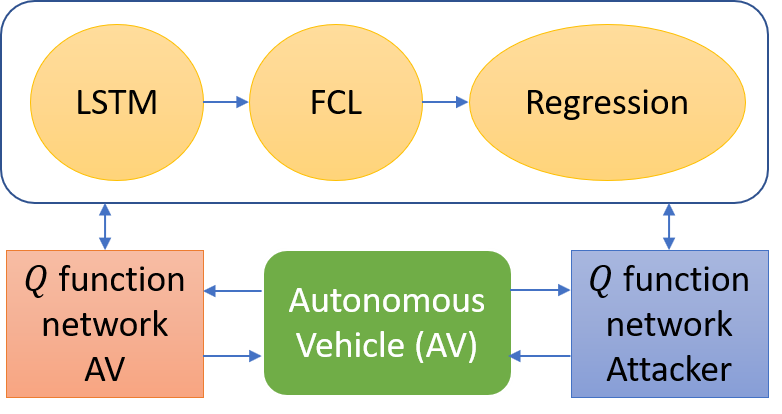}
\caption{The architecture of the adversarial DRL algorithm for robust AV control. A deep neural network (DNN) consisting of a long short term memory (LSTM), a fully connected layer (FCL), and regression is used to learn long-term dependencies within a large data sets, which contain the outcomes of the players' past interactions. The DNN can approximate the Q functions to find optimal actions for players, i.e., AV (defender) and especially attacker, who seeks to inject faulty data to AV sensor readings.}
\label{fig:6}
\end{figure}

Autonomous vehicles (AVs) operating in the \textit{future smart cities} require a robust processing unit of intra-vehicle sensors, including camera, radar, roadside smart sensors, and inter-vehicle beaconing. Such reliance is vulnerable to cyber-physical attacks aiming to get control of AVs by manipulating the sensory data and affecting the reliability of the system, e.g., increasing accident risks or reducing the vehicular flow. Ferdowsi et al. \cite{Ferdowsi2018} examined the scenarios where the attackers manage to interject faulty data to the AV's sensor readings while the AV (the defender) needs to deal with that problem to control AV robustly. Specifically, the \textit{car following model} \cite{Wang2019b} is considered in which the focus is on autonomous control of a car that follows closely another car. The defender aims to learn the leading vehicle's speed based on sensor readings. The attacker's objective is to mislead the following vehicle to a deviation from the optimal safe spacing. The interactions between the attacker and defender are characterized by a game-theoretic problem. The interactive game structure and its DRL solution are diagrammed in Fig. \ref{fig:6}. Instead of directly deriving a solution based on the \textit{mixed-strategy Nash equilibrium} analytics, the authors proposed the use of DRL to solve this dynamic game. Long short term memory (LSTM) \cite{Hochreiter1997} is used to approximate the Q-function for both defending and attacking agents as it can capture the temporal dynamics of the environment.\par

Likewise, Rasheed et al. \cite{Rasheed2020} introduced an adversarial DRL method that integrates LSTM and generative adversarial network (GAN) model to cope with data infusion attacks in autonomous vehicles equipped with 5G communication links. The adversary attempts to inject faulty data to impact safe distance spacing between autonomous vehicles while the autonomous vehicles manage to minimize this deviation. LSTM is used as a generator while a convolutional neural network (CNN) is used as a discriminator to resemble a GAN structure, which is able to capture previous temporal actions of the autonomous vehicle and attacker as well as previous distance deviations. A DRL algorithm is proposed based on these observations to select optimal actions (suitable velocity) for the autonomous vehicle to avoid collisions and accidents. 

Autonomous systems can be vulnerable to inefficiency from various sources such as noises in communication channels, sensor failures, errors in sensor measurement readings, packet errors, and especially cyber attacks. \textit{Deception attack} to autonomous systems is widespread as it is initiated by an adversary whose effort is to inject noises to the communication channels between sensors and the command center. This kind of attack leads to corrupted information being sent to the command center and eventually degrades the system performance. Gupta and Yang \cite{Gupta2018} studied the ways to increase the robustness of autonomous systems by allowing the system to learn using adversarial examples. The problem is formulated as a zero-sum game with the players to be the command center (observer) and the adversary. The inverted pendulum problem from Roboschool \cite{Schulman2017} is used as a simulation environment. The TRPO algorithm is employed to design an observer that can reliably detect adversarial attacks in terms of measurement corruption and automatically mitigate their effects.\par

\subsection{DRL-based Intrusion Detection Systems}
To detect intrusions, security experts conventionally need to observe and examine audit data, e.g., application traces, network traffic flow, and user command data, to differentiate between normal and abnormal behaviors. However, the volume of audit data surges rapidly when the network size is enlarged. This makes manual detection difficult or even impossible. An intrusion detection system (IDS) is a software or hardware platform installed on host computers or network equipment to detect and report to the administrator abnormal or malicious activities by analysing the audit data. An intrusion detection and prevention system may be able to take appropriate actions immediately to reduce impacts of the malevolent activities. 

Depending on different types of audit data, IDSs are grouped into two categories: host-based or network-based IDS. \textit{Host-based IDS} typically observes and analyses the host computer's log files or settings to discover anomalous behaviors. \textit{Network-based IDS} relies on a sniffer to collect transmitting packets in the network and examines the traffic data for intrusion detection. Host-based systems normally lack cross-platform support and implementing them requires knowledge of the host operating systems and configurations. Network-based systems aim to monitor traffic over specific network segments, and they are independent of the operating systems and more portable than host-based systems. Implementing network-based systems is thus easier and they offer more monitoring power than host-based systems. However, a network-based IDS may have difficulty in handling heavy traffic and high-speed networks because it must examine every packet passing through its network segment.

Regardless of the IDS type, two common detection methods are used: signature-based and anomaly-based detection. \textit{Signature detection} involves the storage of patterns of known attacks and comparing characteristics of possible attacks to those in the database. \textit{Anomaly detection} observes the normal behaviors of the system and alerts the administrator if any activities are found deviated from normality, for instance, the unexpected increase of traffic rate, i.e., number of IP packets per second. Machine learning techniques, including unsupervised clustering and supervised classification methods, have been used widely to build adaptive IDSs \cite{Abubakar2017, Jose2018, Roshan2018, Dey2019, Papamartzivanos2019}. These methods, e.g., neural networks \cite{Haider2016}, k-nearest neighbors \cite{Haider2016, Deshpande2018}, support vector machine (SVM) \cite{Nobakht2016, Haider2016}, random forest \cite{Resende2018} and recently deep learning \cite{Kim2016a, Chawla2018, Hassan2020}, however normally rely on fixed features of existing cyber attacks so that they are deficient in detecting new or deformed attacks. The lack of prompt responses to dynamic intrusions also leads to ineffective solutions produced by unsupervised or supervised techniques. In this regard, RL methods have been demonstrated effectively in various IDS applications \cite{Janagam2018}. The following subsections review the use of DRL methods in both host-based and network-based IDSs.

\subsubsection{Host-based IDS}
As the volume of audit data and complexity of intrusion behaviors increase, adaptive intrusion detection models demonstrate limited effectiveness because they can only handle temporally isolated labeled or unlabeled data. In practice, many complex intrusions comprise temporal sequences of dynamic behaviors. Xu and Xie \cite{Xu2005} proposed an RL-based IDS that can handle this problem. System call trace data are fed into a \textit{Markov reward process} whose state value can be used to detect abnormal temporal behaviors of host processes. The intrusion detection problem is thus converted to a state value prediction task of the Markov chains. The linear \textit{temporal difference (TD) RL} algorithm \cite{Sutton1988} is used as the state value prediction model where its outcomes are compared with a predetermined threshold to distinguish normal traces and attack traces. Instead of using the errors between real values and estimated ones, TD learning algorithm uses the differences between successive approximations to update the state value function. Experimental results obtained from using system call trace data show the dominance of the proposed RL-based IDS in terms of higher accuracy and lower computational costs compared to SVM, hidden Markov model, and other machine learning or data mining methods. The proposed method based on the linear basis functions, however, has a shortcoming when sequential intrusion behaviors are highly nonlinear. Therefore, a kernel-based RL approach using least-squares TD \cite{Xu2006} was suggested for intrusion detection in \cite{Xu2007b, Xu2010}. Relying on the kernel methods, the generalization capability of TD RL is enhanced, especially in high-dimensional and nonlinear feature spaces. The kernel least-squares TD algorithm is, therefore, able to predict anomaly probabilities accurately, which contributes to improving the detection performance of IDS, especially when dealing with multi-stage cyber attacks. 

\subsubsection{Network-based IDS}
Deokar and Hazarnis \cite{Deokar2012} pointed out the drawbacks of both anomaly-based and signature-based detection methods. On the one hand, \textit{anomaly detection} has a high false alarm rate because it may categorize activities which users rarely perform as an anomaly. On the other hand, \textit{signature detection} cannot discover new types of attacks as it uses a database of patterns of well-known attacks. The authors, therefore, proposed an IDS that can identify known and unknown attacks effectively by combining features of both anomaly and signature detection through the use of log files. The proposed IDS is based on a collaboration of RL methods, association rule learning, and log correlation techniques. RL gives a reward (or penalty) to the system when it selects log files that contain (or do not contain) anomalies or any signs of attack. This procedure enables the system to choose more appropriate log files in searching for traces of attack.\par

One of the most difficult challenges in the current Internet is dealing with the \textit{distributed denial-of-service} (DDoS) threat, which is a DoS attack but has the distributed nature, occurring with a large traffic volume, and compromising a large number of hosts. Malialis and Kudenko \cite{Malialis2014, Malialis2015} initially introduced the multiagent router throttling method based on the SARSA algorithm \cite{Sutton1998} to address the DDoS attacks by learning multiple agents to rate-limit or throttle traffic towards a victim server. That method, however, has a limited capability in terms of scalability. They therefore further proposed the coordinated team learning design to the original multiagent router throttling based on the divide-and-conquer paradigm to eliminate the mentioned drawback. The proposed approach integrates three mechanisms, namely, task decomposition, hierarchical team-based communication, and team rewards, involving multiple defensive nodes across different locations to coordinately stop or reduce the flood of DDoS attacks. A network emulator is developed based on the work of Yau et al. \cite{Yau2005} to evaluate throttling approaches. Simulation results show that the resilience and adaptability of the proposed method are superior to its competing methods, i.e., baseline router throttling and additive-increase/multiplicative-decrease throttling algorithms \cite{Yau2005}, in various scenarios with different attack dynamics. The scalability of the proposed method is successfully experimented with up to 100 RL agents, which has a great potential to be deployed in a large internet service provider network.\par

\begin{figure}[tp]
\centering
\includegraphics[width=0.4\textwidth]{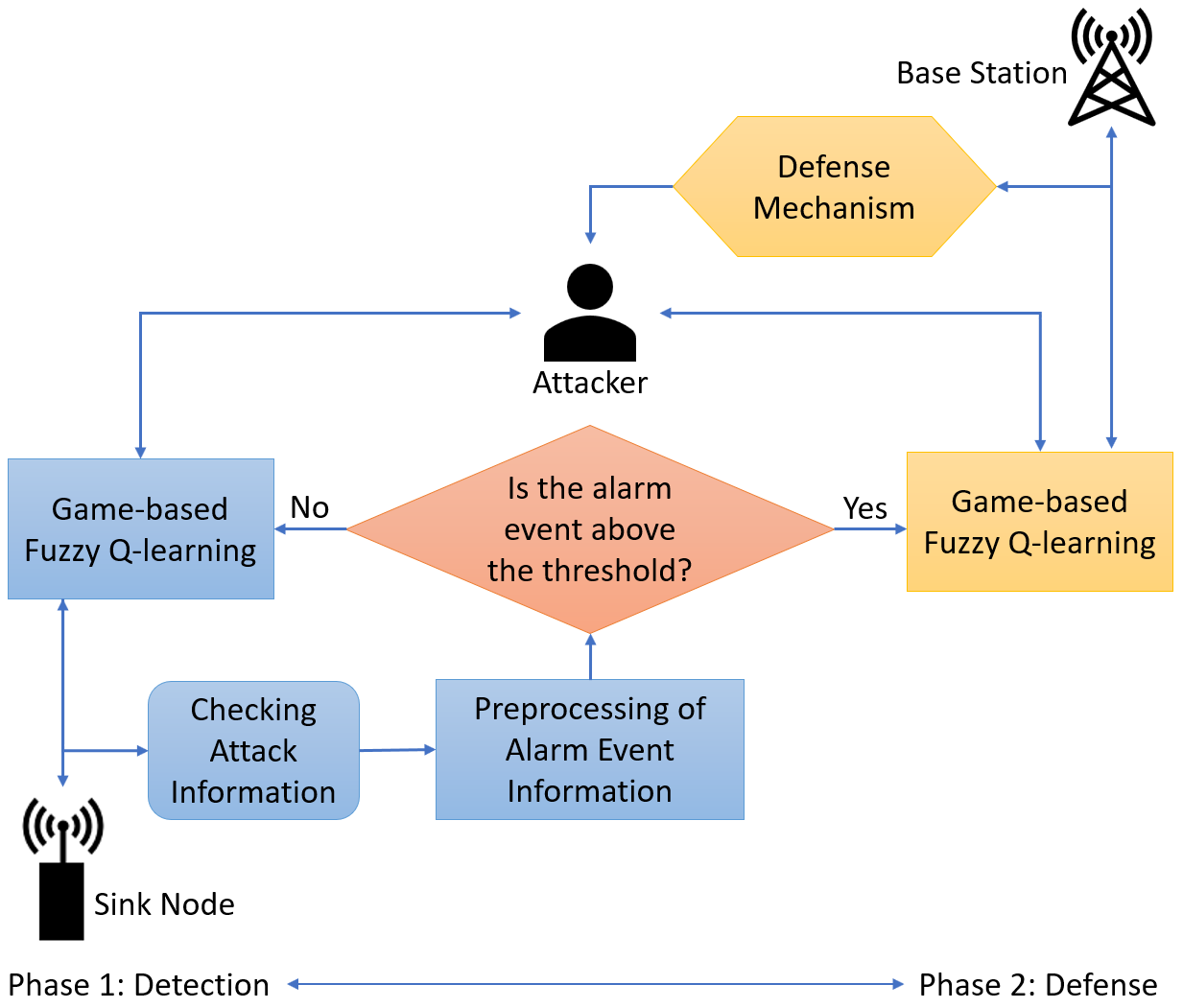}
\caption{Two-phase intrusion detection and prevention system based on a game theory approach and fuzzy Q-learning. In Phase 1, the sink node uses fuzzy Q-learning to detect anomalies caused by the attacker to victim nodes. The malicious information is preprocessed and checked against a threshold by the sink node before passing to Phase 2 where the base station also employs fuzzy Q-learning to select optimal defense actions.}
\label{fig:7}
\end{figure}

Alternatively, Bhosale et al. \cite{Bhosale2014} proposed a multiagent intelligent system \cite{Herrero2009} using RL and influence diagram \cite{Detwarasiti2005} to enable quick responses against the complex attacks. Each agent learns its policy based on local database and information received from other agents, i.e., decisions and events. Shamshirband et al. \cite{Shamshirband2014}, on the other hand, introduced an intrusion detection and prevention system for wireless sensor networks (WSNs) based on a game theory approach and employed a fuzzy Q-learning algorithm \cite{Munoz2013, Shamshirband2013} to obtain optimal policies for the players. Sink nodes, a base station, and an attacker constitute a three-player game where sink nodes and base station are coordinated to derive a defense strategy against the DDoS attacks, particularly in the application layer. The IDS detects future attacks based on the fuzzy Q-learning algorithm that takes part in two phases: detection and defense (Fig. \ref{fig:7}). The game commences when the attacker sends an overwhelming volume of flooding packets beyond a specific threshold as a DDoS attack to a victim node in the WSN. Using the low energy adaptive clustering hierarchy (LEACH), which is a prominent WSN protocol \cite{Varshney2018}, the performance of the proposed method is evaluated and compared with that of existing soft computing methods. The results show the efficacy and viability of the proposed method in terms of detection accuracy, energy consumption and network lifetime.\par

In another approach, Caminero et al. \cite{Caminero2019} proposed a model, namely adversarial environment using RL, to incorporate the RL theory in implementing a classifier for network intrusion detection. A simulated environment is created where random samples drawn from a labelled network intrusion dataset are treated as RL states. The adversarial strategy is employed to deal with unbalanced datasets as it helps to avoid training bias via an oversampling mechanism and thus decrease classification errors for under-represented classes. Likewise, a study in \cite{Lopez2020} applied DRL methods such as DQN, double DQN (DDQN), policy gradient and actor-critic models for network intrusion detection. With several adjustments and adaptations, DRL algorithms can be used as a supervised approach to classifying labelled intrusion data. DRL policy network is simple and fast, which is suitable for online learning and rapid responses in modern data networks with evolving environments. Results obtained on two intrusion detection datasets show DDQN is the best algorithm among the four employed DRL algorithms. DDQN’s performance is equivalent and even better than many traditional machine learning methods in some cases. Recently, Saeed et al. \cite{Saeed2020} examined the existing multiagent IDS architectures, including several approaches that utilized RL algorithms. The adaptation capability of RL methods can help IDS to respond effectively to changes in the environments. However, optimal solution is not guaranteed because convergence of a multiagent system is hard to obtain.

\subsection{DRL-based Game Theory for Cyber Security}
Traditional cyber security methods such as firewall, anti-virus software, or intrusion detection are normally passive, unilateral, and lagging behind dynamic attacks. Cyberspace involves various cyber components, and thus, reliable cyber security requires the consideration of interactions among these components. Specifically, the security policy applied to a component has a certain impact on the decisions taken by other components. Therefore, the decision space increases considerably, with many what-if scenarios when the system is large. Game theory has been demonstrated effectively in solving such large-scale problems because it can examine many scenarios to derive the best policy for each player \cite{Roy2010, Shiva2010, Ramachandran2016, Wang2016b, Zhu2018b}. The utility or payoff of a game player depends not only on its actions but also on other players' activities. In other words, the efficacy of cyber defending strategies must take into account the attacker's strategies and other network users' behaviors. Game theory can model the conflict and cooperation between intelligent decision makers, resembling activities of cyber security problems, which involve attackers and defenders. This resemblance has enabled game theory to mathematically describe and analyze the complex behaviors of multiple competitive mechanisms. In the following, we present game theoretic models involving multiple DRL agents that characterize cyber security problems in different attacking scenarios, including jamming, spoofing, malware, and attacks in adversarial environments.

\subsubsection{Jamming attacks}
\begin{figure}[tp]
\centering
\includegraphics[width=0.43\textwidth]{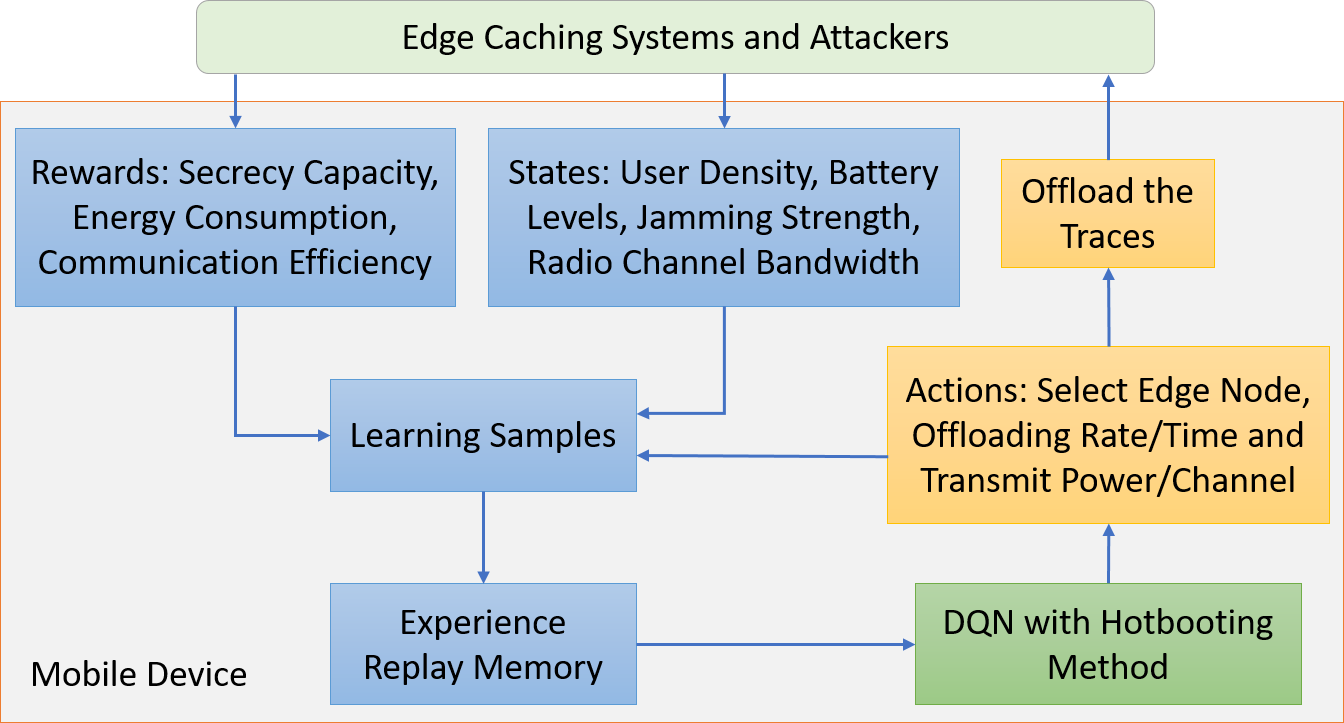}
\caption{Secure offloading method in MEC based on DQN with hotbooting technique. The DQN agent' actions are to find optimal parameters such as offloading rate, power, and channel for the mobile device to offload the traces to the edge node accordingly. The attackers may deploy jamming, spoofing, DoS, or smart attacks to disrupt this process. By interacting with the edge caching systems, the agent can evaluate the reward of the previous action and obtain new state, enabling it to select the next optimal action.}
\label{fig:10}
\end{figure}

Jamming attacks can be considered as a special case of DoS attacks, which are defined as any event that diminishes or eradicates a network's capacity to execute its expected function \cite{Mpitziopoulos2009, Hu2018, Boche2019}. Jamming is a serious attack in networking and has attracted a great interest of researchers who used machine learning or especially RL to address this problem, e.g., \cite{Wu2012, Singh2012, Gwon2013, Conley2013, Dabcevic2014, Slimeni2015, Xiao20181}. The recent development of deep learning has facilitated the use of DRL for jamming handling or mitigation. Xiao et al. \cite{Xiao2018b} studied security challenges of the MEC systems and proposed an RL-based solution to provide secure offloading to the edge nodes against jamming attacks. MEC is a technique that allows cloud computing functions to take place at the edge nodes of a cellular network or generally of any network. This technology helps to decrease network traffic, reduce overhead and latency when users request to access contents that have been cached in the edges closer to the cellular customer. MEC systems, however, are vulnerable to cyber attacks because they are physically located closer to users and attackers with less secure protocols compared to cloud servers or database center. In \cite{Xiao2018b}, the RL methodology is used to select the defense levels and important parameters such as offloading rate and time, transmit channel and power. As the network state space is large, the authors proposed the use of DQN to handle high-dimensional data, as illustrated in Fig. \ref{fig:10}. DQN uses a CNN to approximate the Q-function that requires high computational complexity and memory. To mitigate this disadvantage a transfer learning method named hotbooting technique is used. The \textit{hotbooting method} helps to initialize weights of CNN more efficiently by using experiences that have been learned in similar circumstances. This reduces the learning time by avoiding random explorations at the start of each episode. Simulation results demonstrate that the proposed method is effective in terms of enhancing the security and user privacy of MEC systems and it can protect the systems in confronting with different types of smart attacks with low overhead.\par

On the other hand, Aref et al. \cite{Aref2017} introduced a multiagent RL method to deal with anti-jamming communications in \textit{wideband autonomous cognitive radios} (WACRs). WACRs are advanced radios that can sense the states of the radio frequency spectrum and network, and autonomously optimize its operating mode corresponding to the perceived states. Cognitive communication protocols, however, may struggle when there are unintentional interferences or malicious users who attempt to interrupt reliable communications by deliberate jamming. Each radio's effort is to occupy the available common wideband spectrum as much as possible and avoid sweeping signal of a jammer that affects the entire spectrum band. The multiagent RL approach proposed in \cite{Aref2017} learns an optimal policy for each radio to select appropriate sub-band, aiming to avoid jamming signals and interruptions from other radios. Comparative studies show the significant dominance of the proposed method against a random policy. A drawback of the proposed method is the assumption that the jammer uses a fixed strategy in responding to the WACRs strategies, although the jammer may be able to perform adaptive jamming with the cognitive radio technology. In \cite{Machuzak2016}, when the current spectrum sub-band is interfered by a jammer, Q-learning is used to optimally select a new sub-band that allows uninterrupted transmission as long as possible. The reward structure of the Q-learning agent is defined as the amount of time that the jammer or interferer takes to interfere with the WACR transmission. Experimental results using the hardware-in-the-loop prototype simulation show that the agent can detect the jamming patterns and successfully learns an optimal sub-band selection policy for jamming avoidance. The obvious drawback of this method is the use of Q-table with a limited number of environment states.\par

The access right to spectrum (or more generally resources) is the main difference between cognitive radio networks (CRNs) and traditional wireless technologies. RL in general or Q-learning has been investigated to produce optimal policy for cognitive radio nodes to interact with their radio-frequency environment \cite{Felice2019}. Attar et al. \cite{Attar2012} examined RL solutions against attacks on both CRN architectures, i.e., infrastructure-based, e.g., the IEEE 802.22 standard, and infrastructure-less, e.g., ad hoc CRN. The adversaries may attempt to manipulate the spectrum sensing process and cause the main sources of security threats in infrastructure-less CRNs. The external adversary node is not part of the CRN, but such attackers can affect the operation of an ad hoc CRN via jamming attacks. In an infrastructure-based CRN, an exogenous attacker can mount incumbent emulation or perform sensor-jamming attacks. The attacker can increase the local false-alarm probability to affect the decision of the IEEE 802.22 base station about the availability of a given band. A jamming attack can have both short-term and long-term effects. Wang et al. \cite{Wang2011} developed a game-theoretic framework to battle against jamming in CRNs where each radio observes the status and quality of available channels and the strategy of jammers to make decisions accordingly. The CRN can learn optimal channel utilization strategy using minimax-Q learning policy \cite{Littman1994}, solving the problems of how many channels to use for data and to control packets along with the channel switching strategy. The performance of minimax-Q learning represented via spectrum-efficient throughput is superior to the myopic learning method, which gives high priority to the immediate payoff and ignores the environment dynamics as well as the attackers' cognitive capability.\par

In CRNs, \textit{secondary users} (SUs) are obliged to avoid disruptions to communications of \textit{primary users} (PUs) and can only gain access to the licensed spectrum when it is not occupied by PUs. Jamming attacks are emergent in CRNs due to the opportunistic access of SUs as well as the appearance of \textit{smart jammers}, which can detect the transmission strategy of SUs. Xiao et al. \cite{Xiao2015a} studied the scenarios where a smart jammer aims to disrupt the SUs rather than PUs. The SUs and jammer, therefore, must sense the channel to check the presence of PUs before making their decisions. The constructed scenarios consist of a secondary source node supported by relay nodes to transmit data packets to secondary receiving nodes. The smart jammer can learn quickly the frequency and transmission power of SUs while SUs do not have full knowledge of the underlying dynamic environment. The interactions between SUs and jammer are modeled as a cooperative transmission power control game, and the optimal strategy for SUs is derived based on the Stackelberg equilibrium \cite{Yang2013}. The aim of SU players is to select appropriate transmission powers to efficiently send data messages in the presence of jamming attacks. The jammer's utility gain is the SUs' loss and vice versa. RL methods, i.e., Q-learning \cite{Watkins1992} and WoLF-PHC \cite{Bowling2002}, are used to model SUs as intelligent agents for coping with the smart jammer. WoLF-PHC stands for the combination of Win or Learn Fast algorithm and policy hill-climbing method. It uses a varying learning rate to foster convergence to the game equilibrium by adjusting the learning speed \cite{Bowling2002}. Simulation results show the improvement in the anti-jamming performance of the proposed method in terms of the \textit{signal to interference plus noise ratio} (SINR). The optimal strategy achieved from the Stackelberg game can minimize the damage created by the jammer in the worst-case scenario.\par

Recently, Han et al. \cite{Han2017} introduced an anti-jamming system for CRNs using the DQN algorithm based on a \textit{frequency-spatial} anti-jamming communication game. The game simulates an environment of numerous jammers that inject jamming signals to disturb the ongoing transmissions of SUs. The SU should not interfere with the communications of PUs and must defeat smart jammers. This communication system is two-dimensional that utilizes both frequency hopping and user mobility. The RL state is the radio environment consisting of PUs, SUs, jammers, and serving base stations/access points. The DQN is used to derive an optimal frequency hopping policy that determines whether the SU should leave an area of heavy jamming or choose a channel to send signals. Experimental results show the superiority of the DQN-based method against the Q-learning based strategy in terms of faster convergence rate, increased SINR, lower cost of defense, and improved utility of the SU. DQN with the core component CNN helps to speed the learning process of the system, which has a large number of frequency channels, compared with the benchmark Q-learning method.\par

To improve the work of Han et al. \cite{Han2017}, Liu et al. \cite{Liu2018} also proposed an anti-jamming communication system using a DRL method but having different and more extensive contributions. Specifically, Liu et al. \cite{Liu2018} used the raw spectrum information with temporal features, known as \textit{spectrum waterfall} \cite{Chen2016} to characterize the environment state rather than using the SINR and PU occupancy as in \cite{Han2017}. Because of this, Liu et al.'s model does not necessitate prior knowledge about the jamming patterns and parameters of the jammer but rather uses the local observation data. This prevents the model from the loss of information and facilitates its adaptability to a dynamic environment. Furthermore, Liu et al.'s work does not assume that the jammer needs to take the same channel-slot transmission structure with the users as in \cite{Han2017}. The recursive CNN is utilized to deal with a complex infinite environment state represented by spectrum waterfall, which has a recursion characteristic. The model is tested using several jamming scenarios, which include the sweeping jamming, comb jamming, dynamic jamming, and intelligent comb jamming. A disadvantage of both Han et al. and Liu et al.'s methods is that they can only derive an optimal policy for one user, which inspires a future research direction focusing on multiple users' scenarios.

\subsubsection{Spoofing attacks}
Spoofing attacks are popular in wireless networks where the attacker claims to be another node using the faked identity such as media access control to gain access to the network illegitimately. This illegal penetration may lead to man-in-the-middle, or DoS attacks \cite{Zeng2010}. Xiao et al. \cite{Xiao2015b, Xiao2016} modeled the interactions between the legitimate receiver and spoofers as a zero-sum authentication game and utilized Q-learning and Dyna-Q \cite{Sutton1990} algorithms to address the spoofing detection problem. The utility of receiver or spoofer is computed based on the Bayesian risk, which is the expected payoff in the spoofing detection. The receiver aims to select the optimal test threshold in PHY-layer spoofing detection while the spoofer needs to select an optimal attacking frequency. To prevent collisions, spoofers are cooperative to attack the receiver. Simulation and experimental results show the improved performance of the proposed methods against the benchmark method with a fixed test threshold. A disadvantage of the proposed approaches is that both action and state spaces are quantized into discrete levels, bounded within a specified interval, which may lead to locally optimal solutions. 

\subsubsection{Malware attacks}
\begin{figure}[htb]
\centering
\includegraphics[width=0.44\textwidth]{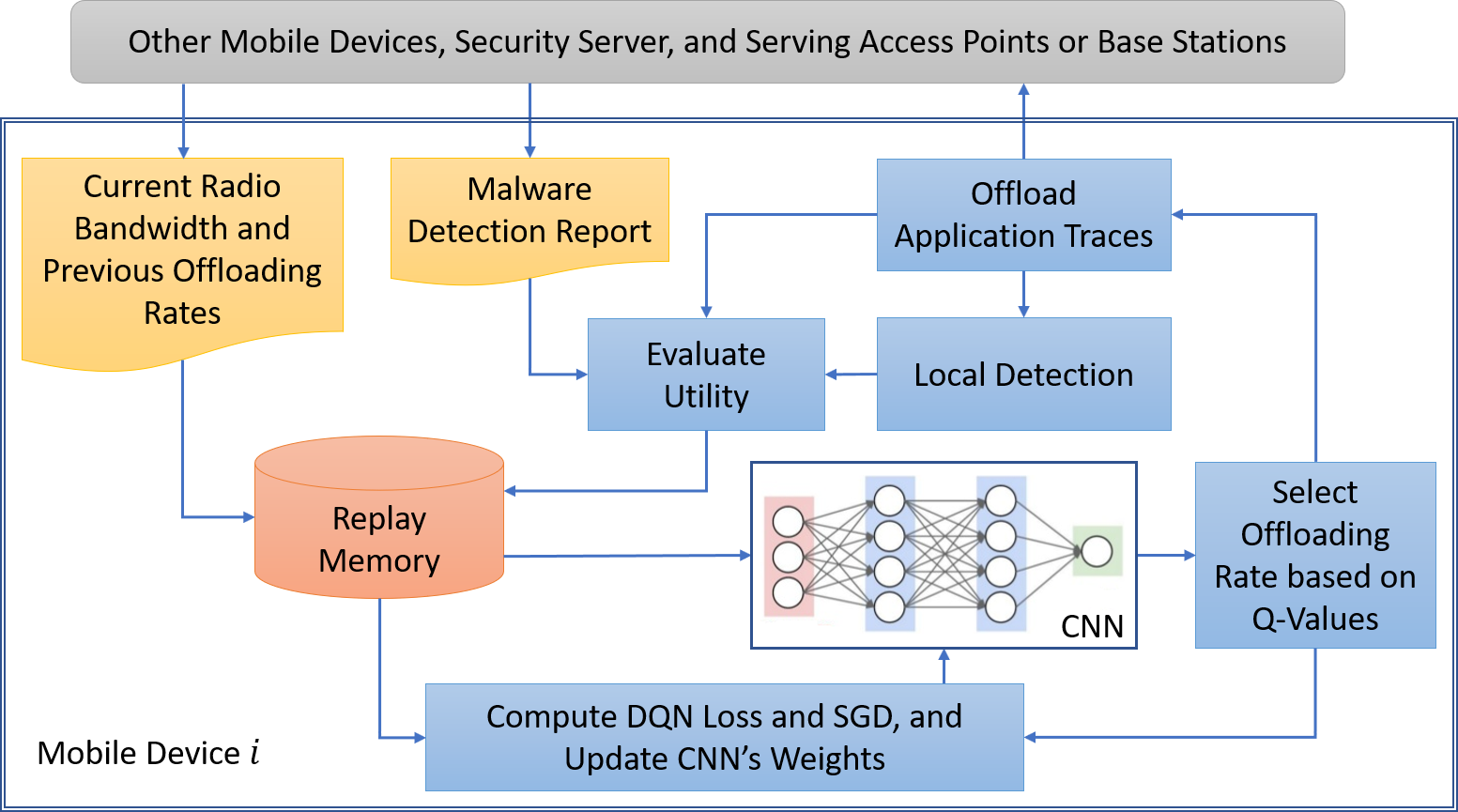}
\caption{Cloud-based malware detection using DQN where the stochastic gradient descent (SGD) method is used to update weights of the CNN. Malicious detection is performed in the cloud server with more powerful computational resources than mobile devices. The DQN agent helps to select optimal task offloading rates for mobile devices to avoid network congestion and detection delay. By observing the network status and evaluating utility based on malware detection report from the server, the agent can formulate states and rewards, which are used to generate a sequence of optimal actions, i.e., dynamic offloading rates.}
\label{fig:8}
\end{figure}

One of the most challenging malwares of mobile devices is the \textit{zero-day attacks}, which exploit publicly unknown security vulnerabilities, and until they are contained or mitigated, hackers might have already caused adverse effects on computer programs, data or networks \cite{Sun2018b, Afek2019}. To avoid such attacks, the traces or log data produced by the applications need to be processed in real time. With limited computational power, battery life and radio bandwidth, mobile devices often offload specific malware detection tasks to security servers at the cloud for processing. The security server with powerful computational resources and more updated malware database can process the tasks quicker, more accurately, and then send a detection report back to mobile devices with less delay. The offloading process is, therefore, a key factor affecting the \textit{cloud-based malware detection} performance. For example, if too many tasks are offloaded to the cloud server, there would be a radio network congestion that can lead to long detection delay. Wan et al. \cite{Wan2017} enhanced the mobile offloading performance by improving the previously proposed game model in \cite{Li2015}. The Q-learning approach used in \cite{Li2015} to select optimal offloading rate suffers the curse of high-dimensionality when the network size is increased, or a large number of feasible offloading rates is available for selection. Wan et al. \cite{Wan2017} thus advocated the use of hotbooting Q-learning and DQN, and showed the performance improvement in terms of malware detection accuracy and speed compared to the standard Q-learning. The cloud-based malware detection approach using DQN for selecting the optimal offloading rate is illustrated in Fig. \ref{fig:8}.\par

\subsubsection{Attacks in adversarial environments}
Traditional networks facilitate the direct communications between client application and server where each network has its switch control that makes the network reconfiguration task time-consuming and inefficient. This method is also disadvantageous because the requested data may need to be retrieved from more than one database involving multiple servers. Software-defined network is a next-generation networking technology as it can reconfigure the network adaptively. With the control being programmable with a global view of the network architecture, SDN can manage and optimize network resources effectively. RL has been demonstrated broadly in the literature as a robust method for SDN controlling, e.g., \cite{Salahuddin2015, Huang2015, Kim2016b, Lin2016, Mestres2017}.\par

Although RL's success in SDN controlling is abundant, the attacker may be able to falsify the defender's training process if it is aware of the network control algorithm in an adversarial environment. To deal with this problem, Han et al. \cite{Han2018} proposed the use of \textit{adversarial RL} to build an \textit{autonomous defense system for SDN}. The attacker selects important nodes in the network to compromise, for example, nodes in the backbone network or the target subnet. By propagating through the network, the attacker attempts to eventually compromise the critical server while the defender prevents the server from being compromised and preserve as many unaffected nodes as possible. To achieve those goals, the RL defender takes four possible actions, consisting of ``isolating", ``patching", ``reconnecting" and ``migrating". Two types of DRL agents are trained to model defenders, i.e., double DQN and A3C, to select appropriate actions given different network states. The reward is characterized based on the status of the critical server, the number of preserved nodes, migration costs, and the validity of the actions taken. That study considered the scenarios where attackers can penetrate the learning process of RL defender by flipping reward signs or manipulating states. These causative attacks poison the defender's training process and cause it to perform sub-optimal actions. The adversarial training approach is applied to reduce the impact of poisoning attacks with its eminent performance demonstrated via several experiments using the popular network emulator Mininet \cite{Lantz2015}.\par

\begin{figure}[tp]
\centering
\includegraphics[width=0.3\textwidth]{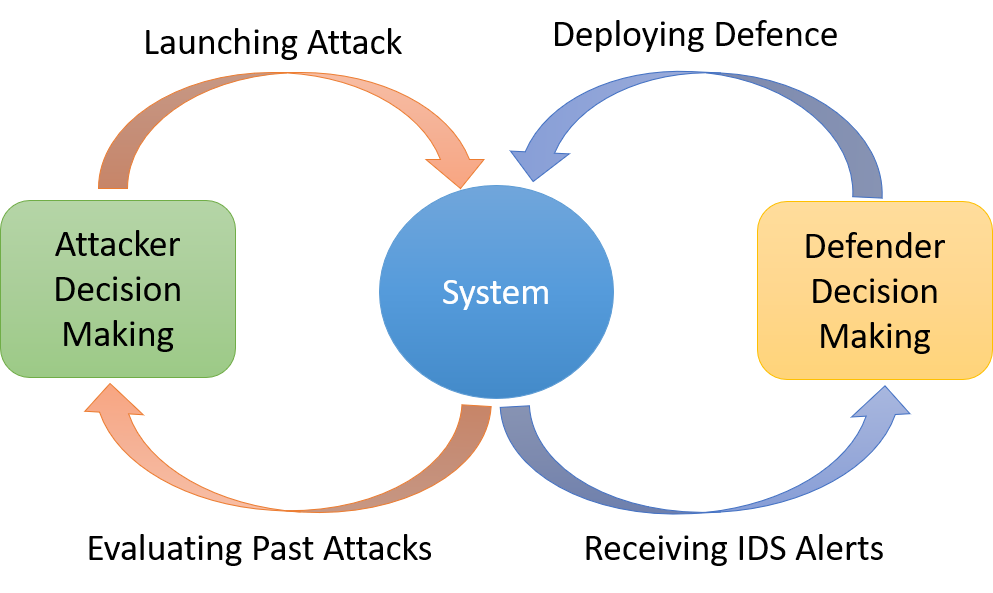}
\caption{The defender and attacker interact via the intrusion detection system (IDS) in an adversarial environment, involving defense and attack cycles. Using these two cycles, a defender and an attacker can repeatedly change their defense and attack strategies. This model can be used to study defense strategies for different classes of attacks such as buffer over-read attacks \cite{Wang2015b} and code reuse attacks \cite{Luo2018}.}
\label{fig:9}
\end{figure}

In an adversarial environment, the defender may not know the private details of the attacker such as the type of attack, attacking target, frequency, and location. Therefore, the defender, for example, may allocate substantial resources to protect an asset that is not a target of the attacker. The defender needs to dynamically reconfigure defense strategies to increase the complexity and cost for the intruder. Zhu et al. \cite{Zhu2014} introduced a model where the defender and attacker can repeatedly change the defense and attack strategies. The defender has no prior knowledge about the attacker, such as launched attacks and attacking policies. However, it is aware of the attacker classes and can access to the system utilities, which are jointly contributed by the defense and attack activities. Two interactive RL methods are proposed for cyber defenses in \cite{Zhu2014}, namely adaptive RL and robust RL. The adaptive RL handles attacks that have a diminishing exploration rate (\textit{non-persistent attacker}) while the robust RL deals with intruders who have a constant exploration rate (\textit{persistent attacker}). The interactions between defender and attacker are illustrated via the attack and defense cycles as in Fig. \ref{fig:9}. The attackers and defenders do not take actions simultaneously but asynchronously. On the attack cycle, the attacker evaluates previous attacks before launching a new attack if necessary. On the defense cycle, after receiving an alert, the defender carries out a meta-analysis on the latest attacks and calculates the corresponding utility before deploying a new defense if needed. An advantage of this system model is that it does not assume any underlying model for the attacker but instead treats attack strategies as black boxes.\par

Alternatively, Elderman et al. \cite{Elderman2017} simulated cyber security problems in networking as a stochastic Markov game with two agents, one attacker, and one defender, with incomplete information and partial observability. The attacker does not know the network topology but attempts to reach and get access to the location that contains a valuable asset. The defender knows the internal network but does not see the attack types or position of intruders. This is a challenging cyber security game because a player needs to adapt its strategy to defeat the unobservable opponents \cite{Chung2016}. Different algorithms, e.g., Monte Carlo learning, Q-learning, and neural networks are used to learn both defender and attacker. Simulation results show that Monte Carlo learning with the softmax exploration is the best method for learning both attacking and defending strategies. Neural network algorithms have a limited adversarial learning capability, and thus they are outperformed by Q-learning and Monte Carlo learning techniques. This simulation has a disadvantage that simplifies the real-world cyber security problem into a game of only two players with only one asset. In real practice, there can be multiple hackers simultaneously penetrating a server that holds valuable data. Also, a network may contain useful data in different locations instead of in a single location as simulated.

\section{Discussions and Future Research Directions}

%TABLE 2
%\begin{center} % remove to get rid of unnecessary space
\begin{table*}[hbt]
\caption{Summary of typical DRL applications in cyber security}
\label{tb:2}
{\scriptsize
%{\fontsize{6.5}{6.5} \selectfont
\hfill{}
\begin{tabular}{|p{2cm}|p{3cm}|p{1.5cm}|p{3cm}|p{3cm}|p{3cm}|}
\hline
Applications & Goals/Objectives & Algorithms & States & Actions & Rewards\\
\hline
Robustness-guided falsification of CPS \cite{Akazaki2018}
& Find falsifying inputs (counterexamples) for CPS
& Double DQN and A3C
& Defined as the output of the system.
& Choose the next input value from a set of piecewise-constant input signals.
& Characterized by a function of past-dependent life-long property, output signal and time.\\
\hline
Security and safety in autonomous vehicle systems \cite{Ferdowsi2018}
& Maximize the robustness of AV dynamics control to cyber-physical attacks that inject faulty data to sensor readings.
& Q-learning with LSTM
& AV's own position and speed along with distance and speed of some nearby objects, e.g., the leading AV.
& Take appropriate speeds to maintain safe spacing between AVs.
& Using a utility function that takes into account the deviation from the optimal safe spacing.\\
\hline
Increasing robustness of the autonomous system against adversarial attacks \cite{Gupta2018}
& Devise filtering schemes to detect corrupted measurements (deception attack) and mitigate the effects of adversarial errors.
& TRPO
& Characterized by sensor measurements and actuation noises.
& Determine which estimation rule to use to generate an estimated state from a corrupted state.
& Defined via a function that takes state features as inputs.\\
\hline
Secure offloading in mobile edge caching \cite{Xiao2018b}
& Learn a policy for a mobile device to securely offload data to edge nodes against jamming and smart attacks.
& DQN with hotbooting transfer learning technique.
& Represented via a combination of user density, battery levels, jamming strength, and radio channel bandwidth.
& Agent's actions include choosing an edge node, selecting offloading rate and time, transmit power and channel.
& Computed based on secrecy capacity, energy consumption, and communication efficiency.\\
\hline
Anti-jamming communication scheme for CRN \cite{Han2017}
& Derive an optimal frequency hopping policy for CRN SUs to defeat smart jammers based on a frequency-spatial anti-jamming game.
& DQN that employs CNN
& Consist of presence status of PUs and SINR information at time $t-1$ received from serving base station or access point.
& SUs take action to leave a geographical area of heavy jamming obstructed by smart jammers or choose a frequency channel to send signals.
& Represented via a utility function based on SINR and transmission cost.\\
\hline
Anti-jamming communication method \cite{Liu2018}, improving the previous work in \cite{Han2017}
& Propose a smart anti-jamming scheme similar to \cite{Han2017} with two main differences: spectrum waterfall is used as the state, and jammers can have different channel-slot transmission structure with users.
& DQN with recursive CNN due to recursion characteristic of spectrum waterfall.
& Using temporal and spectral information, i.e., spectrum waterfall containing both frequency and time domain information of the network environment.
& Agent's action is to select a discretized transmission frequency from a predefined set.
& Defined by a function involving SINR-based transmission rate and cost for frequency switching.\\
\hline
Spoofing detection in wireless networks \cite{Xiao2015b, Xiao2016}
& Select the optimal authentication threshold.
& Q-learning and Dyna-Q
& Include false alarm rate and missed detection rate of the spoofing detection system at time $t-1$
& Action set includes the choices of different discrete levels of the authentication thresholds bounded within a specified interval.
& Using a utility function calculated based on the Bayesian risk, which is the expected payoff in spoofing detection.\\
\hline
Mobile offloading for cloud-based malware detection \cite{Wan2017}, improving the previous work in \cite{Li2015}
& Improve malware detection accuracy and speed.
& Hotbooting Q-learning and DQN.
& Consist of current radio bandwidth and previous offloading rates of other devices.
& Select optimal offloading rate level for each mobile device.
& Represented by a utility function calculated based on the detection accuracy, response speed, and transmission cost.\\
\hline
Autonomous defense in SDN \cite{Han2018}
& Tackle the poisoning attacks that manipulate states or flip reward signals during the training process of RL-based defense agents.
& Double DQN and A3C
& Represented by an array of zeros and ones showing the state of the network (whether a node is compromised or a link is switched on/off). Array length is equal to several nodes plus several links.
& Attackers learn to select a node to compromise while a defender can take four actions: isolate, patch, reconnect and migrate to protect server and preserve as many nodes as possible.
& Modelled based on the status of the critical server, number of preserved nodes, migration cost and the validity of actions taken.\\
\hline
Secure mobile crowdsensing (MCS) system \cite{Xiao2017}
& Optimize payment policy to improve the sensing performance against faked sensing attacks by formulating a Stackelberg game.
& DQN
& Consist of the previous sensing quality and the payment policy.
& Select the server's optimal payment vector to smartphone users.
& Using a utility function that involves the total payment to users and the benefit of server from sensing reports of different accuracy levels.\\
\hline
Automated URL-based phishing detection \cite{Chatterjee2019}
& Detect malicious websites (URLs)
& DQN
& Characterized by the vector space representation of website features such as HTTPS protocols, having IP address, prefix or suffix in URLs.
& Select either 0 or 1, corresponding to a benign or phishing URL.
& Based on the classification action, the reward equates to 1 or -1 if the URL is classified correctly or incorrectly.\\
\hline
\end{tabular}
} %for small
\hfill{}
\end{table*}
%\end{center}

DRL has emerged over recent years as one of the most successful methods of designing and creating human or even superhuman AI agents. Many of these successes have relied on the incorporation of DNNs into a framework of traditional RL to address complex and high-dimensional sequential decision-making problems. Applications of DRL algorithms, therefore, have been found in various fields, including IoT and cyber security. Computers and the Internet today play crucial roles in many areas of our lives, e.g., entertainment, communication, transportation, medicine, and even shopping. Lots of our personal information and important data are stored online. Even financial institutions, e.g., banks, mortgage companies, and brokerage firms, run their business online. Therefore, it is essential to have a security plan in place to prevent hackers from accessing our computer systems. This paper has presented a comprehensive survey of DRL methods and their applications to cyber security problems, with notable examples summarized in Table \ref{tb:2}. The adversarial environment of cyber systems has instigated various proposals of game theory models involving multiple DRL agents. We found that this kind of application occupies a major proportion of papers in the literature relating to DRL for cyber security problems.

\subsection{Challenges and future work on applying DRL for CPS security solutions}
An emerging area is the use of DRL for security solutions for cyber-physical systems \cite{Wei2020, Liu2020c}. The large-scale and complex nature of CPSs, e.g., in environmental monitoring networks, electrical smart grid systems, transportation management network, and cyber manufacturing management system, require security solutions to be responsive and accurate. This has been addressed by various DRL approaches, e.g., TRPO algorithm \cite{Gupta2018}, LSTM-Q-learning \cite{Ferdowsi2018}, double DQN, and A3C \cite{Akazaki2018}. One of the great challenges in implementing DRL algorithms for CPS security solutions is the lack of realistic CPS simulations. For example, the work in \cite{Akazaki2018} had to use the Matlab/Simulink CPS modelling and embed it into the OpenAI Gym environment. This implementation is expensive in terms of computational time due to the overhead caused by integrating the Matlab/Simulink CPS simulation in the OpenAI Gym library. More proper simulations of CPS models embedded directly in DRL-enabled environments are thus encouraged in a future work. Another common challenge in applying DRL algorithms is to transfer the trained policies from simulations to real-world environments. While simulations are cheap and safe for training DRL agents, the reality gap due to modelling impreciseness and errors make the transfer challenging. This is more critical for CPS modelling because of the complexity, dynamics and large scale of CPS. A research in this direction, i.e., sim-to-real transfer for DRL-based security solutions for CPS, is worth investigating as it can help to reduce time, costs and increase safety during training process of DRL agents, and eventually reduce costly mistakes when executing in real-world environments.
 
\subsection{Challenges and future work on applying DRL for IDS}
Although there have been a large number of applications of traditional RL methods for IDSs, there has been a small amount of work on DRL algorithms for this kind of application. This is probably because the integration of deep learning and RL methods has just been sustained very recently. The complexity and dynamics of intrusion detection problems are expected to be solved effectively by DRL methods, which combine the powerful representation learning and function approximation capabilities of deep learning and the optimal sequential decision-making capability of traditional RL. Applying DRL for IDS requires simulated or real intrusion environments for training agents interactively. This is a great challenge because using real environments for training is costly while simulated environments may be far from reality. Most of the existing studies on DRL for intrusion detection relied on game-based settings (e.g., Fig. \ref{fig:7}) or labelled intrusion datasets. For example, the work in \cite{Lopez2020} used two datasets of labelled intrusion samples and adjusted the DRL machinery for it to work on these datasets in a supervised learning manner. This kind of application lacks a live environment and proper interactions between DRL agents and the environment. There is thus a gap for future research on creating more realistic environments, which are able to respond in real time to actions of the DRL agents and facilitate the full exploitation of the DRL's capabilities to solve complex and sophisticated cyber intrusion detection problems. Furthermore, as host-based and network-based IDSs have both advantages and disadvantages, combining these systems could be a logical approach. DRL-based solutions for this kind of integrated system would be another interesting future study.

\subsection{Exploring capabilities of model-based DRL methods}
Most DRL algorithms used for cyber defense so far are model-free methods, which are sample inefficient as they require a large quantity of training data. These data are difficult to obtain in real cyber security practice. Researchers generally utilize simulators to validate their proposed approaches, but these simulators often do not characterize the complexity and dynamics of real cyber space of the IoT systems fully. Model-based DRL methods are more appropriate than model-free methods when training data are limitedly available because, with model-based DRL, it can be easy to collect data in a scalable way. Exploration of model-based DRL methods or the integration of model-based and model-free methods for cyber defense is thus an interesting future study. For example, function approximators can be used to learn a proxy model of the actual high-dimensional and possibly partial observable environment \cite{Oh2015, Mathieu2015, Nagabandi2018}, which can be then employed to deploy planning algorithms, e.g., Monte-Carlo tree search techniques \cite{Browne2012}, to derive optimal actions. Alternatively, model-based and model-free combination approaches, such as model-free policy with planning capabilities \cite{Tamar2016, Pascanu2017} or model-based lookahead search \cite{Silver2016}, can be used as they aggregate advantages of both methods.
On the other hand, current literature on applications of DRL to cyber security often limits at discretizing the action space, which restricts the full capability of the DRL solutions to real-world problems. An example is the application of DRL for selecting optimal mobile offloading rates in \cite{Li2015, Wan2017} where the action space has been discretized although a small change of the rate would primarily affect the performance of the cloud-based malware detection system. Investigation of methods that can deal with continuous action spaces in cyber environments, e.g., policy gradient and actor-critic algorithms, is another encouraging research direction.\par

\subsection{Training DRL in adversarial cyber environments}
AI can help defend against cyber attacks but can also facilitate dangerous attacks, i.e., \textit{offensive AI}. Hackers can take advantages of AI to make attacks smarter and more sophisticated to bypass detection methods to penetrate computer systems or networks. For example, hackers may employ algorithms to observe normal behaviors of users and employ the users' patterns to develop untraceable attacking strategies. Machine learning-based systems can mimic humans to craft convincing fake messages that are utilized to conduct large-scale phishing attacks. Likewise, by creating highly realistic fake video or audio messages based on AI advances (i.e., \textit{deepfakes} \cite{Nguyen2019e}), hackers can spread false news in elections or manipulate financial markets \cite{Giles2019}. Alternatively, attackers can poison the data pool used for training deep learning methods (i.e., \textit{machine learning poisoning}) or attackers can manipulate the states or policies, falsify part of the reward signals in RL to trick the agent into taking sub-optimal actions, resulting in the agent being compromised \cite{Behzadan2017}. These kinds of attacks are difficult to prevent, detect, and fight against as they are part of a battle between AI systems. \textit{Adversarial machine learning}, especially supervised methods, have been used extensively in cyber security \cite{Duddu2018} but very few studies have been found on using adversarial RL \cite{Chen2019}. Adversarial DRL or DRL algorithms trained in various adversarial cyber environments are worth comprehensive investigations as they can be a solution to battle against the increasingly complex offensive AI systems \cite{Ilahi2020, Tong2020, Sun2020}.\par

\subsection{Human-machine teaming with human-on-the-loop models}
With the support of AI systems, cyber security experts no longer examine a huge volume of attack data manually to detect and defend against cyber attacks. This has many advantages because the security teams alone cannot sustain the volume. AI-enabled defense strategies can be automated and deployed rapidly and efficiently but these systems alone cannot issue creative responses when new threats are introduced. Moreover, human adversaries are always behind the cybercrime or cyber warfare. Therefore, there is a critical need for human intellect teamed with machines for cyber defenses. The traditional \textit{human-in-the-loop} model for human-machine integration struggles to adapt quickly with cyber defense system because autonomous agent carries out part of the task and need to halt to wait for human's responses before completing the task. The modern \textit{human-on-the-loop} model would be a solution for a future human-machine teaming cyber security system. This model allows agents to autonomously perform the task whilst humans can monitor and intervene operations of agents only when necessary. How to integrate human knowledge into DRL algorithms \cite{Nguyen2018c} under the human-on-the-loop model for cyber defense is an interesting research question.\par

\subsection{Exploring capabilities of multiagent DRL methods}
As hackers utilized more and more sophisticated and large-scale approach to attack computer systems and networks, the defense strategies need to be more intelligent and large-scale as well. Multiagent DRL is a research direction that can be explored to tackle this problem. Game theory models for cyber security reviewed in this paper have involved multiple agents but they are restricted at a couple of attackers and defenders with limited communication, cooperation and coordination among the agents. These aspects of multiagent DRL need to be investigated thoroughly in cyber security problems to enable an effective large-scale defense plan. Challenges of multiagent DRL itself then need to be addressed such as non-stationarity, partial observability, and efficient multiagent training schemes \cite{Nguyen2018d}. On the other hand, the RL methodology has been applied to deal with various cyber attacks, e.g. jamming, spoofing, false data injection, malware, DoS, DDoS, brute force, Heartbleed, botnet, web attack, and infiltration attack \cite{Xiao2017, Liu20181, Xu2018, Yao2019, Li20191, Chatterjee2019}. However, recently emerged or new types of attacks have been largely unaddressed. One of these new types is the \textit{bit-and-piece DDoS attack}. This attack injects small junk into legitimate traffic of over a large number of IP addresses so that it can bypass many detection methods as there is so little of it per address. Another emerging attack, for instance, is attacking from the computing cloud to breach systems of companies who manage IT systems for other firms or host other firms' data on their servers. Alternatively, hackers can use quantum physics-based powerful computers to crack encryption algorithms that are currently used to protect various types of invaluable data \cite{Giles2019}. Consequently, a future study on addressing these new types of attacks is encouraged.

\ifCLASSOPTIONcaptionsoff
  \newpage
\fi



%\bibliographystyle{IEEEtran}

\begin{thebibliography}{111}
\bibitem{Kakalou2017} I. Kakalou, K. E. Psannis, P. Krawiec, and R. Badea, ``Cognitive radio network and network service chaining toward 5G: challenges and requirements," \textit{IEEE Communications Magazine}, vol. 55, no. 11, pp. 145-151, 2017.

\bibitem{Huang2020} Y. Huang, S. Li, C. Li, Y. T. Hou, and W. Lou, ``A deep reinforcement learning-based approach to dynamic eMBB/URLLC multiplexing in 5G NR," \textit{IEEE Internet of Things Journal}, vol. 7, no. 7, pp. 6439-6456, 2020.

\bibitem{Wang2020} P. Wang, L. T. Yang, X. Nie, Z. Ren, J. Li, and L. Kuang, ``Data-driven software defined network attack detection: state-of-the-art and perspectives," \textit{Information Sciences}, vol. 513, pp. 65-83, 2020.

\bibitem{Botta2016} A. Botta, W. De Donato, V. Persico, and A. Pescapé, ``Integration of cloud computing and Internet of Things: a survey," \textit{Future Generation Computer Systems}, vol. 56, pp. 684-700, 2016.

\bibitem{Krestinskaya2020} O. Krestinskaya, A. P. James, and L. O. Chua, ``Neuromemristive circuits for edge computing: a review," \textit{IEEE Transactions on Neural Networks and Learning Systems}, vol. 31, no. 1, pp. 4-23, 2020.

\bibitem{Abbas2018} N. Abbas, Y. Zhang, A. Taherkordi, and T. Skeie, ``Mobile edge computing: a survey," \textit{IEEE Internet of Things Journal}, vol. 5, no. 1, pp. 450-465, 2018.

\bibitem{Dastjerdi2016} A. V. Dastjerdi, and R. Buyya, ``Fog computing: Helping the Internet of Things realize its potential," \textit{Computer}, vol. 49, no. 8, pp. 112-116, 2016.

\bibitem{Geluvaraj2019} B. Geluvaraj, P. M. Satwik, and T. A. Kumar, ``The future of cybersecurity: major role of artificial intelligence, machine learning, and deep learning in cyberspace," in \textit{International Conference on Computer Networks and Communication Technologies}, 2019, pp. 739-747.

\bibitem{Buczak2016} A. L. Buczak, and E. Guven, ``A survey of data mining and machine learning methods for cyber security intrusion detection," \textit{IEEE Communications Surveys and Tutorials}, vol. 18, no. 2, pp. 1153-1176, 2016.

\bibitem{Apruzzese2018} G. Apruzzese, M. Colajanni, L. Ferretti, A. Guido, and M. Marchetti, ``On the effectiveness of machine and deep learning for cyber security," in \textit{International Conference on Cyber Conflict (CyCon)}, 2018, pp. 371-390.

\bibitem{Xin2018} Y. Xin, L. Kong, Z. Liu, Y. Chen, Y. Li, H. Zhu, M. Gao, H. Hou, and C. Wang, ``Machine learning and deep learning methods for cybersecurity," \textit{IEEE Access}, vol. 6, pp. 35365-35381, 2018.

\bibitem{Milosevic2017} N. Milosevic, A. Dehghantanha, and K. K. R. Choo, ``Machine learning aided Android malware classification," \textit{Computers and Electrical Engineering}, vol. 61, pp. 266-274, 2017.

\bibitem{Mohammed2018} R. Mohammed Harun Babu, R. Vinayakumar, and K. P. Soman, ``A short review on applications of deep learning for cyber security," \textit{arXiv preprint} arXiv:1812.06292, 2018.

\bibitem{Berman2019} D. S. Berman, A. L. Buczak, J. S. Chavis, and C. L. Corbett, ``A survey of deep learning methods for cyber security," \textit{Information}, vol. 10, no. 4, pp. 122, 2019.

\bibitem{Paul2020} S. Paul, Z. Ni, and C. Mu, ``A learning-based solution for an adversarial repeated game in cyber-physical power systems," \textit{IEEE Transactions on Neural Networks and Learning Systems}, DOI: 10.1109/TNNLS.2019.2955857, 2020.

\bibitem{Ding2018} D. Ding, Q. L. Han, Y. Xiang, X. Ge, and X. M. Zhang, ``A survey on security control and attack detection for industrial cyber-physical systems," \textit{Neurocomputing}, vol. 275, pp. 1674-1683, 2018.

\bibitem{Wu2019} M. Wu, Z. Song, and Y. B. Moon, ``Detecting cyber-physical attacks in CyberManufacturing systems with machine learning methods," \textit{Journal of Intelligent Manufacturing}, vol. 30, no. 3, pp. 1111-1123, 2019.

\bibitem{Xiao2018a} L. Xiao, X. Wan, X. Lu, Y. Zhang, and D. Wu, ``IoT security techniques based on machine learning," \textit{arXiv preprint} arXiv:1801.06275, 2018.

\bibitem{Sharma2011} A. Sharma, Z. Kalbarczyk, J. Barlow, and R. Iyer, ``Analysis of security data from a large computing organization," in \textit{Dependable Systems and Networks (DSN), IEEE/IFIP 41st International Conference on}, 2011, pp. 506-517.

\bibitem{Nguyen2017} N. D. Nguyen, T. Nguyen, and S. Nahavandi, ``System design perspective for human-level agents using deep reinforcement learning: A survey," \textit{IEEE Access}, vol. 5, pp. 27091-27102, 2017.

\bibitem{Sui2020} Z. Sui, Z. Pu, J. Yi, and S. Wu, ``Formation control with collision avoidance through deep reinforcement learning using model-guided demonstration," \textit{IEEE Transactions on Neural Networks and Learning Systems}, DOI: 10.1109/TNNLS.2020.3004893, 2020.

\bibitem{Tsantekidis2020} A. Tsantekidis, N. Passalis, A. S. Toufa, K. Saitas-Zarkias, S. Chairistanidis, and A. Tefas, ``Price trailing for financial trading using deep reinforcement learning," \textit{IEEE Transactions on Neural Networks and Learning Systems}, DOI: 10.1109/TNNLS.2020.2997523, 2020.

\bibitem{Nguyen2018b} T. T. Nguyen, ``A multi-objective deep reinforcement learning framework," \textit{arXiv preprint} arXiv:1803.02965, 2018.

\bibitem{Wang2020b} X. Wang, Y. Gu, Y. Cheng, A. Liu, and C. P. Chen, ``Approximate policy-based accelerated deep reinforcement learning," \textit{IEEE Transactions on Neural Networks and Learning Systems}, vol. 31, no. 6, pp. 1820-1830, 2020.

\bibitem{Ling2015} M. H. Ling, K. L. A. Yau, J. Qadir, G. S. Poh, and Q. Ni, ``Application of reinforcement learning for security enhancement in cognitive radio networks," \textit{Applied Soft Computing}, vol. 37, pp. 809-829, 2015.

\bibitem{Wang2019a} Y. Wang, Z. Ye, P. Wan, and J. Zhao, ``A survey of dynamic spectrum allocation based on reinforcement learning algorithms in cognitive radio networks," \textit{Artificial Intelligence Review}, vol. 51, no. 3, pp. 493-506, 2019.

\bibitem{Lu2020} X. Lu, L. Xiao, T. Xu, Y. Zhao, Y. Tang, and W. Zhuang, ``Reinforcement learning based PHY authentication for VANETs," \textit{IEEE Transactions on Vehicular Technology}, vol. 69, no. 3, pp. 3068-3079, 2020.

\bibitem{Alauthman2020} M. Alauthman, N. Aslam, M. Al-Kasassbeh, S. Khan, A. Al-Qerem, and K. K. R. Choo, ``An efficient reinforcement learning-based botnet detection approach," \textit{Journal of Network and Computer Applications}, vol. 150, pp. 102479, 2020.

\bibitem{Mnih2015} V. Mnih, K. Kavukcuoglu, D. Silver, A. A. Rusu, J. Veness, M. G. Bellemare, ... and S. Petersen, ``Human-level control through deep reinforcement learning," \textit{Nature}, vol. 518, no. 7540, pp. 529-533, 2015.

\bibitem{Nguyen2018a} N. D. Nguyen, S. Nahavandi, and T. Nguyen, ``A human mixed strategy approach to deep reinforcement learning," in \textit{2018 IEEE International Conference on Systems, Man, and Cybernetics (SMC)}, 2018, pp. 4023-4028.

\bibitem{Silver2016} D. Silver, A. Huang, C. J. Maddison, A. Guez, L. Sifre, G. Van Den Driessche, ... and S. Dieleman, ``Mastering the game of Go with deep neural networks and tree search," \textit{Nature}, vol. 529, no. 7587, pp. 484-489, 2016.

\bibitem{Silver2017} D. Silver, J. Schrittwieser, K. Simonyan, I. Antonoglou, A. Huang, A. Guez, ... and Y. Chen, ``Mastering the game of Go without human knowledge," \textit{Nature}, vol. 550, no. 7676, pp. 354-359, 2017.

\bibitem{Vinyals2017} O. Vinyals, T. Ewalds, S. Bartunov, P. Georgiev, A. S. Vezhnevets, M. Yeo, ... and J. Quan, ``StarCraft II: A new challenge for reinforcement learning," \textit{arXiv preprint} arXiv:1708.04782, 2017.

\bibitem{Sun2018a} P. Sun, X. Sun, L. Han, J. Xiong, Q. Wang, B. Li, ... and T. Zhang, ``TStarBots: Defeating the cheating level builtin AI in StarCraft II in the full game," \textit{arXiv preprint} arXiv:1809.07193, 2018.

\bibitem{Pang2018} Z. J. Pang, R. Z. Liu, Z. Y. Meng, Y. Zhang, Y. Yu, and T. Lu, ``On reinforcement learning for full-length game of StarCraft," \textit{arXiv preprint} arXiv:1809.09095, 2018.

\bibitem{Zambaldi2018} V. Zambaldi, D. Raposo, A. Santoro, V. Bapst, Y. Li, I. Babuschkin, ... and M. Shanahan, ``Relational deep reinforcement learning," \textit{arXiv preprint} arXiv:1806.01830, 2018.

\bibitem{Jaderberg2018} M. Jaderberg, W. M. Czarnecki, I. Dunning, L. Marris, G. Lever, A. G. Castaneda, ... and N. Sonnerat, ``Human-level performance in first-person multiplayer games with population-based deep reinforcement learning," \textit{arXiv preprint} arXiv:1807.01281, 2018.

\bibitem{OpenAI2019} OpenAI, ``OpenAI Five," [Online]. Available: https://openai.com/five/, 2019, March 1.

\bibitem{Gu2017} S. Gu, E. Holly, T. Lillicrap, and S. Levine, ``Deep reinforcement learning for robotic manipulation with asynchronous off-policy updates," in \textit{2017 IEEE International Conference on Robotics and Automation (ICRA)}, 2017, pp. 3389-3396.

\bibitem{Isele2018} D. Isele, R. Rahimi, A. Cosgun, K. Subramanian, and K. Fujimura, ``Navigating occluded intersections with autonomous vehicles using deep reinforcement learning," in \textit{2018 IEEE International Conference on Robotics and Automation (ICRA)}, 2018, pp. 2034-2039.

\bibitem{Nguyen2019a} T. T. Nguyen, N. D. Nguyen, F. Bello, and S. Nahavandi, ``A new tensioning method using deep reinforcement learning for surgical pattern cutting," in \textit{2019 IEEE International Conference on Industrial Technology (ICIT)}, DOI: 10.1109/ICIT.2019.8755235, 2019.

\bibitem{Nguyen2019b} N. D. Nguyen, T. Nguyen, S. Nahavandi, A. Bhatti, and G. Guest, ``Manipulating soft tissues by deep reinforcement learning for autonomous robotic surgery," in \textit{2019 IEEE International Systems Conference (SysCon)}, DOI: 10.1109/SYSCON.2019.8836924, 2019.

\bibitem{Keneshloo2020} Y. Keneshloo, T. Shi, N. Ramakrishnan, and C. K. Reddy, ``Deep reinforcement learning for sequence-to-sequence models," \textit{IEEE Transactions on Neural Networks and Learning Systems}, vol. 31, no. 7, pp. 2469-2489, 2020.

\bibitem{Mahmud2018} M. Mahmud, M. S. Kaiser, A. Hussain, and S. Vassanelli, ``Applications of deep learning and reinforcement learning to biological data," \textit{IEEE Transactions on Neural Networks and Learning Systems}, vol. 29, no. 6, pp. 2063-2079, 2018.

\bibitem{Popova2018} M. Popova, O. Isayev, and A. Tropsha, ``Deep reinforcement learning for de novo drug design," \textit{Science Advances}, vol. 4, no. 7, pp. eaap7885, 2018.

\bibitem{He2017} Y. He, F. R. Yu, N. Zhao, V. C. Leung, and H. Yin, ``Software-defined networks with mobile edge computing and caching for smart cities: A big data deep reinforcement learning approach," \textit{IEEE Communications Magazine}, vol. 55, no. 12, pp. 31-37, 2017.

\bibitem{Hasselt2016} H. V. Hasselt, A. Guez, and D. Silver, ``Deep reinforcement learning with double Q-learning," in \textit{The Thirtieth AAAI Conference on Artificial Intelligence}, 2016, pp. 2094-2100.

\bibitem{Wang2016a} Z. Wang, T. Schaul, M. Hessel, H. Hasselt, M. Lanctot, and N. Freitas, ``Dueling network architectures for deep reinforcement learning," in \textit{International Conference on Machine Learning}, 2016, pp. 1995-2003.

\bibitem{Zhu2018a} H. Zhu, Y. Cao, W. Wang, T. Jiang, and S. Jin, ``Deep reinforcement learning for mobile edge caching: Review, new features, and open issues," \textit{IEEE Network}, vol. 32, no. 6, pp. 50-57, 2018.

\bibitem{Mnih2016} V. Mnih, A. P. Badia, M. Mirza, A. Graves, T. Lillicrap, T. Harley, ... and K. Kavukcuoglu, ``Asynchronous methods for deep reinforcement learning," in \textit{International Conference on Machine Learning}, 2016, pp. 1928-1937.

\bibitem{Zhang20171} Y. Zhang, J. Yao, and H. Guan, ``Intelligent cloud resource management with deep reinforcement learning," \textit{IEEE Cloud Computing}, vol. 4, no. 6, pp. 60-69, 2017.

\bibitem{Zhu20181} J. Zhu, Y. Song, D. Jiang, and H. Song, ``A new deep-Q-learning-based transmission scheduling mechanism for the cognitive Internet of Things," \textit{IEEE Internet of Things Journal}, vol. 5, no. 4, pp. 2375-2385, 2017.

\bibitem{Shafin2020} R. Shafin, H. Chen, Y. H. Nam, S. Hur, J. Park, J. Zhang, ... and L. Liu, ``Self-tuning sectorization: deep reinforcement learning meets broadcast beam optimization," \textit{IEEE Transactions on Wireless Communications}, vol. 19, no. 6, pp. 4038-4053, 2020.

\bibitem{Zhang2018} D. Zhang, X. Han, and C. Deng, ``Review on the research and practice of deep learning and reinforcement learning in smart grids," \textit{CSEE Journal of Power and Energy Systems}, vol. 4, no. 3, pp. 362-370, 2018.

\bibitem{He20181} X. He, K. Wang, H. Huang, T. Miyazaki, Y. Wang, and S. Guo, ``Green resource allocation based on deep reinforcement learning in content-centric IoT," \textit{IEEE Transactions on Emerging Topics in Computing}, DOI: 10.1109/TETC.2018.2805718, 2018.

\bibitem{He20182} Y. He, C. Liang, R. Yu, and Z. Han, ``Trust-based social networks with computing, caching and communications: A deep reinforcement learning approach," \textit{IEEE Transactions on Network Science and Engineering}, DOI: 10.1109/TNSE.2018.2865183, 2018.

\bibitem{Luong2019} N. C. Luong, D. T. Hoang, S. Gong, D. Niyato, P. Wang, Y. C. Liang, and D. I. Kim, ``Applications of deep reinforcement learning in communications and networking: A survey," \textit{IEEE Communications Surveys and Tutorials}. DOI: 10.1109/COMST.2019.2916583, 2019.

\bibitem{Dai2019} Y. Dai, D. Xu, S. Maharjan, Z. Chen, Q. He, and Y. Zhang, ``Blockchain and deep reinforcement learning empowered intelligent 5G beyond," \textit{IEEE Network}, vol. 33, no. 3, pp. 10-17, 2019.

\bibitem{Leong2020} A. S. Leong, A. Ramaswamy, D. E. Quevedo, H. Karl, and L. Shi, ``Deep reinforcement learning for wireless sensor scheduling in cyber–physical systems," \textit{Automatica}, vol. 113, pp. 108759, 2020.

\bibitem{Watkins1992} C. J. Watkins, and P. Dayan, ``Q-learning," \textit{Machine Learning}, vol. 8, no. 3-4, pp. 279-292, 1992.

\bibitem{Arulkumaran2017} K. Arulkumaran, M. P. Deisenroth, M. Brundage, and A. A. Bharath, ``Deep reinforcement learning: A brief survey," \textit{IEEE Signal Processing Magazine}, vol. 34, no. 6, pp. 26-38, 2017.

\bibitem{Mnih2013} V. Mnih, K. Kavukcuoglu, D. Silver, A. Graves, I. Antonoglou, D. Wierstra, and M. Riedmiller, ``Playing Atari with deep reinforcement learning," \textit{arXiv preprint} arXiv:1312.5602, 2013.

\bibitem{Schaul2015} T. Schaul, J. Quan, I. Antonoglou, and D. Silver, ``Prioritized experience replay," \textit{arXiv preprint} arXiv:1511.05952, 2015.

\bibitem{Williams1992} R. J. Williams, ``Simple statistical gradient-following algorithms for connectionist reinforcement learning," \textit{Machine Learning}, vol. 8, no. 3-4, pp. 229-256, 1992.

\bibitem{Sutton2000} R. S. Sutton, D. A. McAllester, S. P. Singh, and Y. Mansour, ``Policy gradient methods for reinforcement learning with function approximation," in \textit{Advances in Neural Information Processing Systems}, 2000, pp. 1057-1063.

\bibitem{Schulman2015} J. Schulman, S. Levine, P. Abbeel, M. Jordan, and P. Moritz, ``Trust region policy optimization," in \textit{International Conference on Machine Learning}, 2015, pp. 1889-1897.

\bibitem{Schulman2017} J. Schulman, F. Wolski, P. Dhariwal, A. Radford, and O. Klimov, ``Proximal policy optimization algorithms," \textit{arXiv preprint} arXiv:1707.06347, 2017.

\bibitem{Wu2018} C. Wu, A. Rajeswaran, Y. Duan, V. Kumar, A. M. Bayen, S. Kakade, ... and P. Abbeel, ``Variance reduction for policy gradient with action-dependent factorized baselines," \textit{arXiv preprint} arXiv:1803.07246, 2018.

\bibitem{Lillicrap2015} T. P. Lillicrap, J. J. Hunt, A. Pritzel, N. Heess, T. Erez, Y. Tassa, ... and D. Wierstra, ``Continuous control with deep reinforcement learning," \textit{arXiv preprint} arXiv:1509.02971, 2015.

\bibitem{Barth-Maron2018} G. Barth-Maron, M. W. Hoffman, D. Budden, W. Dabney, D. Horgan, A. Muldal, ... and T. Lillicrap, ``Distributed distributional deterministic policy gradients," \textit{arXiv preprint} arXiv:1804.08617, 2018.

\bibitem{Jaderberg2016} M. Jaderberg, V. Mnih, W. M. Czarnecki, T. Schaul, J. Z. Leibo, D. Silver, and K. Kavukcuoglu, ``Reinforcement learning with unsupervised auxiliary tasks," \textit{arXiv preprint} arXiv:1611.05397, 2016.

\bibitem{Nachum2017} O. Nachum, M. Norouzi, K. Xu, and D. Schuurmans, ``Bridging the gap between value and policy based reinforcement learning," in \textit{Advances in Neural Information Processing Systems}, 2017, pp. 2775-2785.

\bibitem{Lapan2018} M. Lapan, \textit{Deep Reinforcement Learning Hands-On: Apply modern RL methods, with deep Q-networks, value iteration, policy gradients, TRPO, AlphaGo Zero and more}, Packt Publishing Ltd., 2018.

\bibitem{OpenAI1} OpenAI Gym Toolkit Documentation. Classic control: control theory problems from the classic RL literature. Retrieved December 14, 2020, from: https://gym.openai.com/envs/\#classic\_control

\bibitem{OpenAI2} OpenAI Gym Toolkit Documentation. Box2D: continuous control tasks in the Box2D simulator. Retrieved December 14, 2020, from: https://gym.openai.com/envs/\#box2d

\bibitem{Wang2015a} L. Wang, M. Torngren, and M. Onori, ``Current status and advancement of cyber-physical systems in manufacturing," \textit{Journal of Manufacturing Systems}, vol. 37, pp. 517-527, 2015.

\bibitem{Zhang2017} Y. Zhang, M. Qiu, C. W. Tsai, M. M. Hassan, and A. Alamri, ``Health-CPS: Healthcare cyber-physical system assisted by cloud and big data," \textit{IEEE Systems Journal}, vol. 11, no. 1, pp. 88-95, 2017.

\bibitem{Shakeel2018} P. M. Shakeel, S. Baskar, V. S. Dhulipala, S. Mishra, and M. M. Jaber, ``Maintaining security and privacy in health care system using learning based deep-Q-networks," \textit{Journal of Medical Systems}, vol. 42, no. 10, pp. 186, 2018.

\bibitem{Cintuglu2017} M. H. Cintuglu, O. A. Mohammed, K. Akkaya, and A. S. Uluagac, ``A survey on smart grid cyber-physical system testbeds," \textit{IEEE Communications Surveys and Tutorials}, vol. 19, no. 1, pp. 446-464, 2017.

\bibitem{Chen2018} Y. Chen, S. Huang, F. Liu, Z. Wang, and X. Sun, ``Evaluation of reinforcement learning-based false data injection attack to automatic voltage control," \textit{IEEE Transactions on Smart Grid}, vol. 10, no. 2, pp. 2158-2169, 2018.

\bibitem{Ni2019} Z. Ni, and S. Paul, ``A multistage game in smart grid security: A reinforcement learning solution," \textit{IEEE Transactions on Neural Networks and Learning Systems}, DOI: 10.1109/TNNLS.2018.2885530, 2019.

\bibitem{Ferdowsi2020} A. Ferdowsi, A. Eldosouky, and W. Saad, ``Interdependence-aware game-theoretic framework for secure intelligent transportation systems," \textit{IEEE Internet of Things Journal}. DOI: 10.1109/JIOT.2020.3020899, 2020.

\bibitem{Li2018} Y. Li, L. Zhang, H. Zheng, X. He, S. Peeta, T. Zheng, and Y. Li, ``Nonlane-discipline-based car-following model for electric vehicles in transportation-cyber-physical systems," \textit{IEEE Transactions on Intelligent Transportation Systems}, vol. 19, no. 1, pp. 38-47, 2018.

\bibitem{Li2019} C. Li, and M. Qiu, \textit{Reinforcement Learning for Cyber-Physical Systems: with Cybersecurity Case Studies}, CRC Press, 2019.

\bibitem{Feng2017} M. Feng, and H. Xu, ``Deep reinforcement learning based optimal defense for cyber-physical system in presence of unknown cyber attack," in \textit{Computational Intelligence (SSCI), 2017 IEEE Symposium Series on}, 2017, pp. 1-8.

\bibitem{Akazaki2018} T. Akazaki, S. Liu, Y. Yamagata, Y. Duan, and J. Hao, ``Falsification of cyber-physical systems using deep reinforcement learning," in \textit{International Symposium on Formal Methods}, 2018, pp. 456-465.

\bibitem{Abbas2012} H. Abbas, and G. Fainekos, ``Convergence proofs for simulated annealing falsification of safety properties," in \textit{2012 50th Annual Allerton Conference on Communication, Control, and Computing (Allerton)}, 2012, pp. 1594-1601.

\bibitem{Sankaranarayanan2012} S. Sankaranarayanan, and G. Fainekos, ``Falsification of temporal properties of hybrid systems using the cross-entropy method," in \textit{The 15th ACM International Conference on Hybrid Systems: Computation and Control}, 2012, pp. 125-134.

\bibitem{Ferdowsi2018} A. Ferdowsi, U. Challita, W. Saad, and N. B. Mandayam, ``Robust deep reinforcement learning for security and safety in autonomous vehicle systems," in \textit{21st International Conference on Intelligent Transportation Systems (ITSC)}, 2018, pp. 307-312.

\bibitem{Wang2019b} X. Wang, R. Jiang, L. Li, Y. L. Lin, and F. Y. Wang, ``Long memory is important: A test study on deep-learning based car-following model," \textit{Physica A: Statistical Mechanics and its Applications}, vol. 514, pp. 786-795, 2019.

\bibitem{Hochreiter1997} S. Hochreiter, and J. Schmidhuber, ``Long short-term memory," \textit{Neural Computation}, vol. 9, no. 8, pp. 1735-1780, 1997.

\bibitem{Rasheed2020} I. Rasheed, F. Hu, and L. Zhang, ``Deep reinforcement learning approach for autonomous vehicle systems for maintaining security and safety using LSTM-GAN," \textit{Vehicular Communications}, vol. 26, pp. 100266, 2020.

\bibitem{Gupta2018} A. Gupta, and Z. Yang, ``Adversarial reinforcement learning for observer design in autonomous systems under cyber attacks," \textit{arXiv preprint} arXiv:1809.06784, 2018.

\bibitem{Abubakar2017} A. Abubakar, and B. Pranggono, ``Machine learning based intrusion detection system for software defined networks," in \textit{2017 Seventh International Conference on Emerging Security Technologies (EST)}, 2017, pp. 138-143.

\bibitem{Jose2018} S. Jose, D. Malathi, B. Reddy, and D. Jayaseeli, ``A survey on anomaly based host intrusion detection system," in \textit{Journal of Physics: Conference Series}, vol. 1000, no. 1, p. 012049, 2018.

\bibitem{Roshan2018} S. Roshan, Y. Miche, A. Akusok, and A. Lendasse, ``Adaptive and online network intrusion detection system using clustering and extreme learning machines," \textit{Journal of the Franklin Institute}, vol. 355, no. 4, pp. 1752-1779, 2018.

\bibitem{Dey2019} S. Dey, Q. Ye, and S. Sampalli, ``A machine learning based intrusion detection scheme for data fusion in mobile clouds involving heterogeneous client networks," \textit{Information Fusion}, vol. 49, pp. 205-215, 2019.

\bibitem{Papamartzivanos2019} D. Papamartzivanos, F. G. Mármol, and G. Kambourakis, ``Introducing deep learning self-adaptive misuse network intrusion detection systems," \textit{IEEE Access}, vol. 7, pp. 13546-13560, 2019.

\bibitem{Haider2016} W. Haider, G. Creech, Y. Xie, and J. Hu, ``Windows based data sets for evaluation of robustness of host based intrusion detection systems (IDS) to zero-day and stealth attacks," \textit{Future Internet}, vol. 8, no. 3, 29, 2016. 

\bibitem{Deshpande2018} P. Deshpande, S. C. Sharma, S. K. Peddoju, and S. Junaid, ``HIDS: A host based intrusion detection system for cloud computing environment," \textit{International Journal of System Assurance Engineering and Management}, vol. 9, no. 3, pp. 567-576, 2018.

\bibitem{Nobakht2016} M. Nobakht, V. Sivaraman, and R. Boreli, ``A host-based intrusion detection and mitigation framework for smart home IoT using OpenFlow," in \textit{11th International Conference on Availability, Reliability and Security (ARES)}, 2016, pp. 147-156. 

\bibitem{Resende2018} P. A. A. Resende, and A. C. Drummond, ``A survey of random forest based methods for intrusion detection systems," \textit{ACM Computing Surveys (CSUR)}, vol. 51, no. 3, pp. 48, 2018.

\bibitem{Kim2016a} G. Kim, H. Yi, J. Lee, Y. Paek, and S. Yoon, ``LSTM-based system-call language modeling and robust ensemble method for designing host-based intrusion detection systems," \textit{arXiv preprint} arXiv:1611.01726, 2016.

\bibitem{Chawla2018} A. Chawla, B. Lee, S. Fallon, and P. Jacob, ``Host based intrusion detection system with combined CNN/RNN model," in \textit{Joint European Conference on Machine Learning and Knowledge Discovery in Databases}, 2018, pp. 149-158.

\bibitem{Hassan2020} M. M. Hassan, A. Gumaei, A. Alsanad, M. Alrubaian, and G. Fortino, ``A hybrid deep learning model for efficient intrusion detection in big data environment," \textit{Information Sciences}, vol. 513, pp. 386-396, 2020.

\bibitem{Janagam2018} A. Janagam, and S. Hossen, ``Analysis of Network Intrusion Detection System with Machine Learning Algorithms (Deep Reinforcement Learning Algorithm)," Dissertation, Blekinge Institute of Technology, Sweden, 2018.

\bibitem{Xu2005} X. Xu, and T. Xie, ``A reinforcement learning approach for host-based intrusion detection using sequences of system calls," in \textit{International Conference on Intelligent Computing}, 2005, pp. 995-1003.

\bibitem{Sutton1988} R. S. Sutton, ``Learning to predict by the methods of temporal differences," \textit{Machine Learning}, vol. 3, no. 1, pp. 9-44, 1988.

\bibitem{Xu2006} X. Xu, ``A sparse kernel-based least-squares temporal difference algorithm for reinforcement learning," in \textit{International Conference on Natural Computation}, 2006, pp. 47-56.

\bibitem{Xu2007b} X. Xu, and Y. Luo, ``A kernel-based reinforcement learning approach to dynamic behavior modeling of intrusion detection," in \textit{International Symposium on Neural Networks}, 2007, pp. 455-464.

\bibitem{Xu2010} X. Xu, ``Sequential anomaly detection based on temporal-difference learning: Principles, models and case studies," \textit{Applied Soft Computing}, vol. 10, no. 3, pp. 859-867, 2010.

\bibitem{Deokar2012} B. Deokar, and A. Hazarnis, ``Intrusion detection system using log files and reinforcement learning," \textit{International Journal of Computer Applications}, vol. 45, no. 19, pp. 28-35, 2012.

\bibitem{Malialis2014} K. Malialis, ``Distributed Reinforcement Learning for Network Intrusion Response," Doctoral Dissertation, University of York, UK, 2014.

\bibitem{Malialis2015} K. Malialis, and D. Kudenko, ``Distributed response to network intrusions using multiagent reinforcement learning," \textit{Engineering Applications of Artificial Intelligence}, vol. 41, pp. 270-284, 2015.

\bibitem{Sutton1998} R. S. Sutton, A. G. Barto, ``Introduction to Reinforcement Learning," MIT Press Cambridge, MA, USA, 1998.

\bibitem{Yau2005} D. K. Yau, J. C. Lui, F. Liang, and Y. Yam, ``Defending against distributed denial-of-service attacks with max-min fair server-centric router throttles," \textit{IEEE/ACM Transactions on Networking}, vol. 13, no. 1, pp. 29-42, 2005.

\bibitem{Bhosale2014} R. Bhosale, S. Mahajan, and P. Kulkarni, ``Cooperative machine learning for intrusion detection system," \textit{International Journal of Scientific and Engineering Research}, vol. 5, no. 1, pp. 1780-1785, 2014. 

\bibitem{Herrero2009} A. Herrero, and E. Corchado, ``Multiagent systems for network intrusion detection: A review," in \textit{Computational Intelligence in Security for Information Systems}, 2009, pp. 143-154.

\bibitem{Detwarasiti2005} A. Detwarasiti, and R. D. Shachter, ``Influence diagrams for team decision analysis," \textit{Decision Analysis}, vol. 2, no. 4, pp. 207-228, 2005.

\bibitem{Shamshirband2014} S. Shamshirband, A. Patel, N. B. Anuar, M. L. M. Kiah, and A. Abraham, ``Cooperative game theoretic approach using fuzzy Q-learning for detecting and preventing intrusions in wireless sensor networks," \textit{Engineering Applications of Artificial Intelligence}, vol. 32, pp. 228-241, 2014.

\bibitem{Munoz2013} P. Muñoz, R. Barco, and I. de la Bandera, ``Optimization of load balancing using fuzzy Q-learning for next generation wireless networks," \textit{Expert Systems with Applications}, vol. 40, no. 4, pp. 984-994, 2013.

\bibitem{Shamshirband2013} S. Shamshirband, N. B. Anuar, M. L. M. Kiah, and A. Patel, ``An appraisal and design of a multiagent system based cooperative wireless intrusion detection computational intelligence technique," \textit{Engineering Applications of Artificial Intelligence}, vol. 26, no. 9, pp. 2105-2127, 2013.

\bibitem{Varshney2018} S. Varshney, and R. Kuma, ``Variants of LEACH routing protocol in WSN: A comparative analysis," in \textit{The 8th International Conference on Cloud Computing, Data Science and Engineering (Confluence)}, 2018, pp. 199-204.

\bibitem{Caminero2019} G. Caminero, M. Lopez-Martin, and B. Carro, ``Adversarial environment reinforcement learning algorithm for intrusion detection," \textit{Computer Networks}, vol. 159, pp. 96-109, 2019.

\bibitem{Lopez2020} M. Lopez-Martin, B. Carro, and A. Sanchez-Esguevillas, ``Application of deep reinforcement learning to intrusion detection for supervised problems," \textit{Expert Systems with Applications}, vol. 141, 112963, 2020.

\bibitem{Saeed2020} I. A. Saeed, A. Selamat, M. F. Rohani, O. Krejcar, and J. A. Chaudhry, ``A systematic state-of-the-art analysis of multiagent intrusion detection," \textit{IEEE Access}, vol. 8, pp. 180184-180209, 2020.

\bibitem{Roy2010} S. Roy, C. Ellis, S. Shiva, D. Dasgupta, V. Shandilya, and Q. Wu, ``A survey of game theory as applied to network security," in \textit{43rd Hawaii International Conference on System Sciences}, 2010, pp. 1-10.

\bibitem{Shiva2010} S. Shiva, S. Roy, and D. Dasgupta, ``Game theory for cyber security," in \textit{The Sixth Annual Workshop on Cyber Security and Information Intelligence Research}, 2010, p. 34.

\bibitem{Ramachandran2016} K. Ramachandran, and Z. Stefanova, ``Dynamic game theories in cyber security," in \textit{International Conference of Dynamic Systems and Applications}, 2016, vol. 7, pp. 303-310.

\bibitem{Wang2016b} Y. Wang, Y. Wang, J. Liu, Z. Huang, and P. Xie, ``A survey of game theoretic methods for cyber security," in \textit{IEEE First International Conference on Data Science in Cyberspace (DSC)}, 2016, pp. 631-636.

\bibitem{Zhu2018b} Q. Zhu, and S. Rass, ``Game theory meets network security: A tutorial," in \textit{The 2018 ACM SIGSAC Conference on Computer and Communications Security}, 2018, pp. 2163-2165.

\bibitem{Mpitziopoulos2009} A. Mpitziopoulos, D. Gavalas, C. Konstantopoulos, and G. Pantziou, ``A survey on jamming attacks and countermeasures in WSNs," \textit{IEEE Communications Surveys and Tutorials}, vol. 11, no. 4, pp. 42-56, 2009.

\bibitem{Hu2018} S. Hu, D. Yue, X. Xie, X. Chen, and X. Yin, ``Resilient event-triggered controller synthesis of networked control systems under periodic DoS jamming attacks," \textit{IEEE Transactions on Cybernetics}, DOI: 10.1109/TCYB.2018.2861834, 2018.

\bibitem{Boche2019} H. Boche, and C. Deppe, ``Secure identification under passive eavesdroppers and active jamming attacks," \textit{IEEE Transactions on Information Forensics and Security}, vol. 14, no. 2, pp. 472-485, 2019.

\bibitem{Wu2012} Y. Wu, B. Wang, K. R. Liu, and T. C. Clancy, ``Anti-jamming games in multi-channel cognitive radio networks," \textit{IEEE Journal on Selected Areas in Communications}, vol. 30, no. 1, pp. 4-15, 2012.

\bibitem{Singh2012} S. Singh, and A. Trivedi, ``Anti-jamming in cognitive radio networks using reinforcement learning algorithms," in \textit{Wireless and Optical Communications Networks (WOCN), Ninth International Conference on}, 2012, pp. 1-5.

\bibitem{Gwon2013} Y. Gwon, S. Dastangoo, C. Fossa, and H. T. Kung, ``Competing mobile network game: Embracing antijamming and jamming strategies with reinforcement learning," in \textit{IEEE Conference on Communications and Network Security (CNS)}, 2013, pp. 28-36.

\bibitem{Conley2013} W. G. Conley, and A. J. Miller, ``Cognitive jamming game for dynamically countering ad hoc cognitive radio networks," in \textit{MILCOM 2013-2013 IEEE Military Communications Conference}, 2013, pp. 1176-1182.

\bibitem{Dabcevic2014} K. Dabcevic, A. Betancourt, L. Marcenaro, and C. S. Regazzoni, ``A fictitious play-based game-theoretical approach to alleviating jamming attacks for cognitive radios," in \textit{2014 IEEE International Conference on Acoustics, Speech and Signal Processing (ICASSP)}, 2014, pp. 8158-8162.

\bibitem{Slimeni2015} F. Slimeni, B. Scheers, Z. Chtourou, and V. Le Nir, ``Jamming mitigation in cognitive radio networks using a modified Q-learning algorithm," in \textit{2015 International Conference on Military Communications and Information Systems (ICMCIS)}, 2015, pp. 1-7.

\bibitem{Xiao20181} L. Xiao, X. Lu, D. Xu, Y. Tang, L. Wang, and W. Zhuang, ``UAV relay in VANETs against smart jamming with reinforcement learning," \textit{IEEE Transactions on Vehicular Technology}, vol. 67, no. 5, pp. 4087-4097, 2018.

\bibitem{Xiao2018b} L. Xiao, X. Wan, C. Dai, X. Du, X. Chen, and M. Guizani, ``Security in mobile edge caching with reinforcement learning," \textit{IEEE Wireless Communications}, vol. 25, no. 3, pp. 116-122, 2018.

\bibitem{Aref2017} M. A. Aref, S. K. Jayaweera, and S. Machuzak, ``Multiagent reinforcement learning based cognitive anti-jamming," in \textit{Wireless Communications and Networking Conference (WCNC)}, 2017, pp. 1-6.

\bibitem{Machuzak2016} S. Machuzak, and S. K. Jayaweera, ``Reinforcement learning based anti-jamming with wideband autonomous cognitive radios," in \textit{2016 IEEE/CIC International Conference on Communications in China (ICCC)}, 2016, pp. 1-5.

\bibitem{Felice2019} M.D. Felice, L. Bedogni, L. Bononi, ``Reinforcement learning-based spectrum management for cognitive radio networks: A literature review and case study," in \textit{Handbook of Cognitive Radio}, 2019, pp. 1849-1886.

\bibitem{Attar2012} A. Attar, H. Tang, A. V. Vasilakos, F. R. Yu, and V. C. Leung, ``A survey of security challenges in cognitive radio networks: Solutions and future research directions," \textit{Proceedings of the IEEE}, vol. 100, no. 12, pp. 3172-3186, 2012.

\bibitem{Wang2011} B. Wang, Y. Wu, K. R. Liu, and T. C. Clancy, ``An anti-jamming stochastic game for cognitive radio networks," \textit{IEEE Journal on Selected Areas in Communications}, vol. 29, no. 4, pp. 877-889, 2011.

\bibitem{Littman1994} M. L. Littman, ``Markov games as a framework for multiagent reinforcement learning," in \textit{The 11th International Conference on Machine Learning}, 1994, pp. 157-163.

\bibitem{Xiao2015a} L. Xiao, Y. Li, J. Liu, and Y. Zhao, ``Power control with reinforcement learning in cooperative cognitive radio networks against jamming," \textit{The Journal of Supercomputing}, vol. 71, no. 9, pp. 3237-3257, 2015.

\bibitem{Yang2013} D. Yang, G. Xue, J. Zhang, A. Richa, and X. Fang, ``Coping with a smart jammer in wireless networks: A Stackelberg game approach," \textit{IEEE Transactions on Wireless Communications}, vol. 12, no. 8, pp. 4038-4047, 2013.

\bibitem{Bowling2002} M. Bowling, and M. Veloso, ``Multiagent learning using a variable learning rate," \textit{Artificial Intelligence}, vol. 136, no. 2, pp. 215-250, 2002.

\bibitem{Han2017} G. Han, L. Xiao, and H. V. Poor, ``Two-dimensional anti-jamming communication based on deep reinforcement learning," in \textit{The 42nd IEEE International Conference on Acoustics, Speech and Signal Processing}, 2017, pp. 2087-2091.

\bibitem{Liu2018} X. Liu, Y. Xu, L. Jia, Q. Wu, and A. Anpalagan, ``Anti-jamming communications using spectrum waterfall: A deep reinforcement learning approach," \textit{IEEE Communications Letters}, vol. 22, no. 5, pp. 998-1001, 2018.

\bibitem{Chen2016} W. Chen, and X. Wen, ``Perceptual spectrum waterfall of pattern shape recognition algorithm," in \textit{The 18th International Conference on Advanced Communication Technology (ICACT)}, 2016, pp. 382-389.

\bibitem{Zeng2010} K. Zeng, K. Govindan, and P. Mohapatra, ``Non-cryptographic authentication and identification in wireless networks," \textit{IEEE Wireless Communications}, vol. 17, no. 5, pp. 56-62, 2010.

\bibitem{Xiao2015b} L. Xiao, Y. Li, G. Liu, Q. Li, and W. Zhuang, ``Spoofing detection with reinforcement learning in wireless networks," in \textit{Global Communications Conference (GLOBECOM)}, 2015, pp. 1-5.

\bibitem{Xiao2016} L. Xiao, Y. Li, G. Han, G. Liu, and W. Zhuang, ``PHY-layer spoofing detection with reinforcement learning in wireless networks," \textit{IEEE Transactions on Vehicular Technology}, vol. 65, no. 12, pp. 10037-10047, 2016.

\bibitem{Sutton1990} R. S. Sutton, ``Integrated architecture for learning, planning, and reacting based on approximating dynamic programming," in \textit{The 7th International Conference on Machine Learning}, 1990, pp. 216-224.

\bibitem{Sun2018b} X. Sun, J. Dai, P. Liu, A. Singhal, and J. Yen, ``Using Bayesian networks for probabilistic identification of zero-day attack paths," \textit{IEEE Transactions on Information Forensics and Security}, vol. 13, no. 10, pp. 2506-2521, 2018.

\bibitem{Afek2019} Y. Afek, A. Bremler-Barr, and S. L. Feibish, ``Zero-day signature extraction for high-volume attacks," \textit{IEEE/ACM Transactions on Networking}, DOI: 10.1109/TNET.2019.2899124, 2019.

\bibitem{Wan2017} X. Wan, G. Sheng, Y. Li, L. Xiao, and X. Du, ``Reinforcement learning based mobile offloading for cloud-based malware detection," in \textit{GLOBECOM 2017-IEEE Global Communications Conference}, 2017, pp. 1-6.

\bibitem{Li2015} Y. Li, J. Liu, Q. Li, and L. Xiao, ``Mobile cloud offloading for malware detections with learning," in \textit{IEEE Conference on Computer Communications Workshops}, 2015, pp. 197-201.

\bibitem{Salahuddin2015} M. A. Salahuddin, A. Al-Fuqaha, and M. Guizani, ``Software-defined networking for RSU clouds in support of the internet of vehicles," \textit{IEEE Internet of Things Journal}, vol. 2, no. 2, pp. 133-144, 2015.

\bibitem{Huang2015} R. Huang, X. Chu, J. Zhang, and Y. H. Hu, ``Energy-efficient monitoring in software defined wireless sensor networks using reinforcement learning: A prototype," \textit{International Journal of Distributed Sensor Networks}, vol. 11, no. 10, 360428, 2015.

\bibitem{Kim2016b} S. Kim, J. Son, A. Talukder, and C. S. Hong, ``Congestion prevention mechanism based on Q-leaning for efficient routing in SDN," in \textit{International Conference on Information Networking (ICOIN)}, 2016, pp. 124-128.

\bibitem{Lin2016} S. C. Lin, I. F. Akyildiz, P. Wang, and M. Luo, ``QoS-aware adaptive routing in multi-layer hierarchical software defined networks: A reinforcement learning approach," in \textit{2016 IEEE International Conference on Services Computing (SCC)}, 2016, pp. 25-33.

\bibitem{Mestres2017} A. Mestres, A. Rodriguez-Natal, J. Carner, P. Barlet-Ros, E. Alarcón, M. Solé, ... and G. Estrada, ``Knowledge-defined networking," \textit{ACM SIGCOMM Computer Communication Review}, vol. 47, no. 3, pp. 2-10, 2017.

\bibitem{Han2018} Y. Han, B. I. Rubinstein, T. Abraham, T. Alpcan, O. De Vel, S. Erfani, ... and P. Montague, ``Reinforcement learning for autonomous defence in software-defined networking," in \textit{International Conference on Decision and Game Theory for Security}, 2018, pp. 145-165.

\bibitem{Lantz2015} B. Lantz, and B. O'Connor, ``A mininet-based virtual testbed for distributed SDN development," in \textit{ACM SIGCOMM Computer Communication Review}, vol. 45, no. 4, pp. 365-366, 2015.

\bibitem{Zhu2014} M. Zhu, Z. Hu, and P. Liu, ``Reinforcement learning algorithms for adaptive cyber defense against Heartbleed," in \textit{The First ACM Workshop on Moving Target Defense}, 2014, pp. 51-58.

\bibitem{Wang2015b} J. Wang, M. Zhao, Q. Zeng, D. Wu, and P. Liu, ``Risk assessment of buffer "Heartbleed" over-read vulnerabilities," in \textit{45th Annual IEEE/IFIP International Conference on Dependable Systems and Networks}, 2015, pp. 555-562.

\bibitem{Luo2018} B. Luo, Y. Yang, C. Zhang, Y. Wang, and B. Zhang, ``A survey of code reuse attack and defense," in \textit{International Conference on Intelligent and Interactive Systems and Applications}, 2018, pp. 782-788.

\bibitem{Elderman2017} R. Elderman, L. J. Pater, A. S. Thie, M. M. Drugan, and M. Wiering, ``Adversarial reinforcement learning in a cyber security simulation," in \textit{International Conference on Agents and Artificial Intelligence (ICAART)}, 2017, vol. 2, pp. 559-566.

\bibitem{Chung2016} K. Chung, C. A. Kamhoua, K. A. Kwiat, Z. T. Kalbarczyk, and R. K. Iyer, ``Game theory with learning for cyber security monitoring," in \textit{IEEE 17th International Symposium on High Assurance Systems Engineering (HASE)}, 2016, pp. 1-8.

\bibitem{Wei2020} F. Wei, Z. Wan, and H. He, ``Cyber-attack recovery strategy for smart grid based on deep reinforcement learning," \textit{IEEE Transactions on Smart Grid}, vol. 11, no. 3, pp. 2476-2486, 2020.

\bibitem{Liu2020c} X. R. Liu, J. Ospina, and C. Konstantinou, ``Deep reinforcement learning for cybersecurity assessment of wind integrated power systems," \textit{arXiv preprint} arXiv:2007.03025, 2020.

\bibitem{Oh2015} J. Oh, X. Guo, H. Lee, R. L. Lewis, and S. Singh, ``Action-conditional video prediction using deep networks in atari games," in \textit{Advances in Neural Information Processing Systems}, 2015, pp. 2863-2871.

\bibitem{Mathieu2015} M. Mathieu, C. Couprie, and Y. LeCun, ``Deep multi-scale video prediction beyond mean square error," \textit{arXiv preprint} arXiv:1511.05440, 2015.

\bibitem{Nagabandi2018} A. Nagabandi, G. Kahn, R. S. Fearing, and S. Levine, ``Neural network dynamics for model-based deep reinforcement learning with model-free fine-tuning," in \textit{2018 IEEE International Conference on Robotics and Automation (ICRA)}, 2018, pp. 7559-7566.

\bibitem{Browne2012} C. B. Browne, E. Powley, D. Whitehouse, S. M. Lucas, P. I. Cowling, P. Rohlfshagen, ... and S. Colton, ``A survey of Monte Carlo tree search methods," \textit{IEEE Transactions on Computational Intelligence and AI in Games}, vol. 4, no. 1, pp. 1-43, 2012.

\bibitem{Tamar2016} A. Tamar, Y. Wu, G. Thomas, S. Levine, and P. Abbeel, ``Value iteration networks," in \textit{Advances in Neural Information Processing Systems}, 2016, pp. 2154-2162.

\bibitem{Pascanu2017} R. Pascanu, Y. Li, O. Vinyals, N. Heess, L. Buesing, S. Racanière, ... and P. Battaglia, ``Learning model-based planning from scratch," \textit{arXiv preprint} arXiv:1707.06170, 2017.

\bibitem{Nguyen2019e} T. T. Nguyen, C. M. Nguyen, D. T. Nguyen, D. T. Nguyen, and S. Nahavandi, ``Deep learning for deepfakes creation and detection: a survey," \textit{arXiv preprint} arXiv:1909.11573, 2019.

\bibitem{Giles2019} M. Giles, ``Five emerging cyber-threats to worry about in 2019," \textit{MIT Technology Review}. [Online]. Available at: https://www.technologyreview.com/2019/01/04/66232/five-emerging-cyber-threats-2019/, 2019, January 4.

\bibitem{Behzadan2017} V. Behzadan, and A. Munir, ``Vulnerability of deep reinforcement learning to policy induction attacks," in \textit{International Conference on Machine Learning and Data Mining in Pattern Recognition}, 2017, pp. 262-275.

\bibitem{Duddu2018} V. Duddu, ``A survey of adversarial machine learning in cyber warfare," \textit{Defence Science Journal}, vol. 68, no. 4, pp. 356-366, 2018.

\bibitem{Chen2019} T. Chen, J. Liu, Y. Xiang, W. Niu, E. Tong, and Z. Han, ``Adversarial attack and defense in reinforcement learning-from AI security view," \textit{Cybersecurity}, vol. 2, no. 1, p. 11, 2019.

\bibitem{Ilahi2020} I. Ilahi, M. Usama, J. Qadir, M. U. Janjua, A. Al-Fuqaha, D. T. Hoang, and D. Niyato, ``Challenges and countermeasures for adversarial attacks on deep reinforcement learning," \textit{arXiv preprint} arXiv:2001.09684, 2020.

\bibitem{Tong2020} L. Tong, A. Laszka, C. Yan, N. Zhang, and Y. Vorobeychik, ``Finding needles in a moving haystack: prioritizing alerts with adversarial reinforcement learning," in \textit{Proceedings of the AAAI Conference on Artificial Intelligence}, 2020, vol. 34, no. 1, pp. 946-953.

\bibitem{Sun2020} J. Sun, T. Zhang, X. Xie, L. Ma, Y. Zheng, K. Chen, and Y. Liu, ``Stealthy and efficient adversarial attacks against deep reinforcement learning," \textit{arXiv preprint} arXiv:2005.07099, 2020.

\bibitem{Nguyen2018c} T. Nguyen, N. D. Nguyen, and S. Nahavandi, ``Multi-agent deep reinforcement learning with human strategies," in \textit{2019 IEEE International Conference on Industrial Technology (ICIT)}, 2019, DOI: 10.1109/ICIT.2019.8755032.

\bibitem{Nguyen2018d} T. T. Nguyen, N. D. Nguyen, and S. Nahavandi, ``Deep reinforcement learning for multiagent systems: a review of challenges, solutions, and applications," \textit{IEEE Transactions on Cybernetics}, vol. 50, no. 9, pp. 3826-3839, 2020.

\bibitem{Xiao2017} L. Xiao, Y. Li, G. Han, H. Dai, and H. V. Poor, ``A secure mobile crowdsensing game with deep reinforcement learning," \textit{IEEE Transactions on Information Forensics and Security}, vol. 13, no. 1, pp. 35-47, 2017.

\bibitem{Liu20181} Y. Liu, M. Dong, K. Ota, J. Li, and J. Wu, ``Deep reinforcement learning based smart mitigation of DDoS flooding in software-defined networks," in \textit{IEEE 23rd International Workshop on Computer Aided Modeling and Design of Communication Links and Networks (CAMAD)}, 2018, pp. 1-6.

\bibitem{Xu2018} Y. Xu, G. Ren, J. Chen, X. Zhang, L. Jia, and L. Kong, ``Interference-aware cooperative anti-jamming distributed channel selection in UAV communication networks," \textit{Applied Sciences}, vol. 8, no. 10, 1911, 2018.

\bibitem{Yao2019} F. Yao, and L. Jia, ``A collaborative multiagent reinforcement learning anti-jamming algorithm in wireless networks," \textit{IEEE Wireless Communications Letters}, DOI: 10.1109/LWC.2019.2904486, 2019.

\bibitem{Li20191} Y. Li, X. Wang, D. Liu, Q. Guo, X. Liu, J. Zhang, and Y. Xu, ``On the performance of deep reinforcement learning-based anti-jamming method confronting intelligent jammer," \textit{Applied Sciences}, vol. 9, no. 7, pp. 1361, 2019.

\bibitem{Chatterjee2019} M. Chatterjee, and A. S. Namin, ``Detecting phishing websites through deep reinforcement learning," in \textit{IEEE 43rd Annual Computer Software and Applications Conference (COMPSAC)}, 2019, vol. 2, pp. 227-232.

\end{thebibliography}

\begin{IEEEbiography}[{\includegraphics[width=1in,height=1.25in,clip,keepaspectratio]{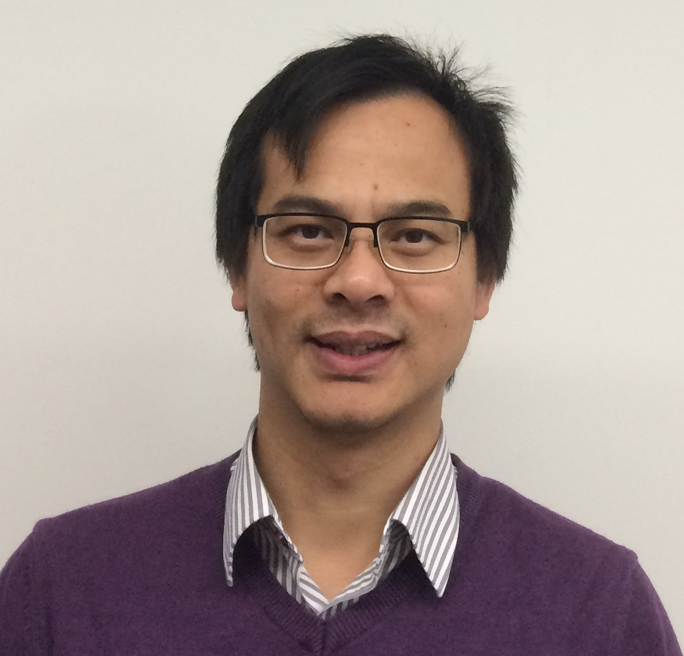}}]{Thanh Thi Nguyen}
was a Visiting Scholar with the Computer Science Department at Stanford University, California, USA in 2015 and the Edge Computing Lab, John A. Paulson School of Engineering and Applied Sciences, Harvard University, Massachusetts, USA in 2019. He received an Alfred Deakin Postdoctoral Research Fellowship in 2016, a European-Pacific Partnership for ICT Expert Exchange Program Award from European Commission in 2018, and an Australia–India Strategic Research Fund Early- and Mid-Career Fellowship Awarded by the Australian Academy of Science in 2020. Dr. Nguyen obtained a PhD in Mathematics and Statistics from Monash University, Australia in 2013 and has expertise in various areas, including artificial intelligence, deep learning, deep reinforcement learning, cyber security, IoT, and data science. He is currently a Senior Lecturer in the School of Information Technology, Deakin University, Victoria, Australia.
\end{IEEEbiography}

\balance

\begin{IEEEbiography}[{\includegraphics[width=1in,height=1.25in,clip,keepaspectratio]{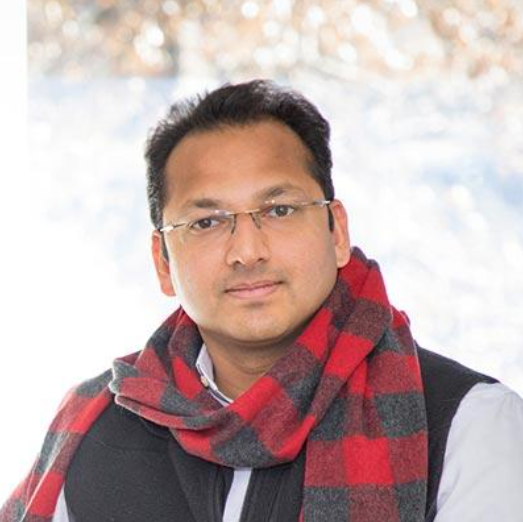}}]{Vijay Janapa Reddi}
completed his PhD in computer science at Harvard University in 2010. He is a recipient of multiple awards, including the National Academy of Engineering (NAE) Gilbreth Lecturer Honor (2016), IEEE TCCA Young Computer Architect Award (2016), Intel Early Career Award (2013), Google Faculty Research Awards (2012, 2013, 2015, 2017), Best Paper at the 2005 International Symposium on Microarchitecture, Best Paper at the 2009 International Symposium on High Performance Computer Architecture, and IEEE's Top Picks in Computer Architecture Awards (2006, 2010, 2011, 2016, 2017). Dr. Reddi is currently an Associate Professor in the John A. Paulson School of Engineering and Applied Sciences at Harvard University where he directs the Edge Computing Lab. His research interests include computer architecture and system-software design, specifically in the context of mobile and edge computing platforms based on machine learning.
\end{IEEEbiography}

\end{document}